\documentclass[sigplan, screen]{acmart}

\usepackage{amsmath}
\usepackage{caption}
\usepackage{subfig}
\usepackage{graphicx}
\usepackage{pifont}
\usepackage{soul}
\usepackage{color,xcolor}
\usepackage{url}

\newcommand{\name}{TensorTEE}

\definecolor{red}{rgb}{1.00,0.00,0.00}
\definecolor{gray}{rgb}{0.5,0.5,0.5}

\copyrightyear{2024}
\acmYear{2024}
\setcopyright{rightsretained}
\acmConference[ASPLOS '24]{29th ACM International Conference on
Architectural Support for Programming Languages and Operating Systems,
Volume 4}{4.27-5.1, 2024}{La Jolla, CA, USA}
\acmBooktitle{29th ACM International Conference on Architectural Support for
Programming Languages and Operating Systems, Volume 4 (ASPLOS '24), April
27-May 1, 2024, La Jolla, CA, USA}
\acmDOI{10.1145/3622781.3674168}
\acmISBN{979-8-4007-0391-1/24/04}

\usepackage[]{hyperref}

\title{
\name: Unifying Heterogeneous TEE Granularity for Efficient Secure Collaborative Tensor Computing
}

\author{
Husheng Han\textsuperscript{1,2,3} \quad Xinyao Zheng\textsuperscript{1,2,3} \quad Yuanbo Wen\textsuperscript{1} 
\quad Yifan Hao\textsuperscript{1}  \quad Erhu Feng\textsuperscript{4,9} \\
Ling Liang\textsuperscript{5} \quad Jianan Mu\textsuperscript{1,2} \quad Xiaqing Li\textsuperscript{1} 
\quad Tianyun Ma\textsuperscript{1,3,6} \quad Pengwei Jin\textsuperscript{1,2,3}  \\ 
Xinkai Song\textsuperscript{1} \quad Zidong Du\textsuperscript{1,8} \quad Qi Guo\textsuperscript{1} 
\quad Xing Hu\textsuperscript{1,7}\footnotemark \\
\small{\textsuperscript{1}\textit{SKLP, Institute of Computing Technology, Chinese Academy of Sciences}  \quad
\textsuperscript{2}\textit{University of Chinese Academy of Sciences} \\
\textsuperscript{3}\textit{Cambricon Technologies} \quad
\textsuperscript{4}\textit{IPADS, Shanghai Jiao Tong University} \quad \textsuperscript{5}\textit{Peking university} \\
\textsuperscript{6}\textit{University of Science and Technology of China} \quad
\textsuperscript{7}\textit{ZGC LAB}\\
\textsuperscript{8}\textit{Shanghai Innovation Center for Processor Technologies, SHIC} \\
\textsuperscript{9}\textit{Engineering Research Center for Domain-specific Operating Systems (MoE)} \\}
\small{
\{ hanhusheng20z, \ zhengxinyao22s, \ wenyuanbo, \ haoyifan, \ mujianan19s, \ jinpengwei20z \}@ict.ac.cn \\
fengerhu1@sjtu.edu.cn \quad lingliang@pku.edu.cn \quad mty21@mail.ustc.edu.cn \\
\{ lixiaqing, \ songxinkai, \ duzidong, \ guoqi, \ huxing \}@ict.ac.cn \\
}
}

\renewcommand{\shortauthors}{Han et al.}

\begin{document}

\begin{abstract}
Heterogeneous collaborative computing with NPU and CPU has received widespread attention due to its substantial performance benefits. To ensure data confidentiality and integrity during computing, Trusted Execution Environments (TEE) is considered a promising solution because of its comparatively lower overhead.
However, existing heterogeneous TEE designs are inefficient for collaborative computing due to fine and different memory granularities between CPU and NPU. 
1) The cacheline granularity of CPU TEE intensifies memory pressure due to its extra memory access, 
and 2) the cacheline granularity MAC of NPU escalates the pressure on the limited memory storage. 
3) Data transfer across heterogeneous enclaves relies on the transit of non-secure regions, resulting in cumbersome re-encryption and scheduling.

To address these issues, we propose ~\name~, a unified tensor-granularity heterogeneous TEE for efficient secure collaborative tensor computing. 
First, we virtually support tensor granularity in CPU TEE to eliminate the off-chip metadata access by detecting and maintaining tensor structures on-chip. 
Second, we propose tensor-granularity MAC management with predictive execution to avoid computational stalls while eliminating off-chip MAC storage and access.
Moreover, based on the unified granularity, we enable direct data transfer without re-encryption and scheduling dilemmas.
Our evaluation is built on enhanced Gem5 and a cycle-accurate NPU simulator. The results show that ~\name~ improves the performance of Large Language Model (LLM) training workloads by 4.0x compared to existing work and incurs only 2.1\% overhead compared to non-secure training, offering a practical security assurance for LLM training.
\end{abstract}

\maketitle 


\let\thefootnote\relax\footnote{* Xing Hu is the corresponding author (huxing@ict.ac.cn).}
\addtocounter{footnote}{-1}\let\thefootnote\svthefootnote

\section{Introduction}
\label{intro}

The heterogeneous collaborative computing with NPU and CPU has demonstrated significant performance and energy improvement in various fields~\cite{ren2021zero,hildebrand2020autotm,huang2020swapadvisor, jin2018layer, peng2020capuchin, ren2021sentinel, rhu2016vdnn, choukse2020buddy}, 
especially when dealing with the challenges posed by the limited DRAM capacity of NPUs, such as LLM training~\cite{ren2021zero}, autonomous driving~\cite{tesla}, and AlphaFold2~\cite{AlphaFold2021},  {hence heterogeneous collaborative architectures emerge as a trend in state-of-the-arts: Nvidia Grace Hopper~\cite{GH200} and AMD Instinct MI300~\cite{MI300}}. 
The improvement stems from the heterogeneous collaborative computing system that can effectively leverage the characteristics of different architectures.
For example, ZeRO-offload~\cite{ren2021zero} of DeepSpeed~\cite{rasley2020deepspeed} offloads forward and backward computations to an NPU, while the CPU manages high-precision weights and optimizer states.
Such a heterogeneous collaborative way effectively alleviates DRAM storage pressure, thus enabling the training of a 10B parameter GPT-2 on a single V100 GPU~\cite{ren2021zero}.

Given that collaborative computing is equally important in terms of computing efficiency and data privacy, the outstanding performance of heterogeneous collaborative computing has increasingly sparked people's pursuit of security.
This trend makes Trusted Execution Environments (TEE), which has been developed for CPU~\cite{costan2016sgx, lee2020keystone, devices2006amd64} and extended to NPU architectures~\cite{volos2018graviton,jang2019hix,costan2016sgx,lee2022tnpu,shrivastava2023securator}, gain significant attention, 
as it can provide secure enclaves with robust privacy guarantees and lower overhead than alternative solutions (such as Fully Homomorphic Encryption and Multiple Party Computation).
Specifically, TEE considers only the on-chip hardware trustworthy, so both CPU/NPU off-chip memory and inter-chip communication require additional protection mechanisms against attacks from OS or physical attacks. In this paper, we focus on discrete NPUs with \textit{off-chip GDDR memory} 
following the same configuration with recent works~\cite{na2021common,hua2022mgx,abdullah2023plutus}.

However, existing heterogeneous TEE designs~\cite{hua2022mgx, shrivastava2023securator}, with different memory protection granularities between CPU and NPU, incur significant performance overhead and are not feasible for collaborative computing. 
They face efficiency issues from three aspects: 
1) The cacheline granularity of Version Number (VN) employed in CPU TEE generates substantial additional memory accesses for VN and Merkle Tree traversal, intensifies memory pressure in memory-intensive workloads, and results in significant overhead.
2) Although VN is tensor-wise managed in NPUs, their cacheline granularity MAC management brings extra storage overhead, further escalating the strain on the already constrained memory storage. 
3) Due to the granularity differences in heterogeneous TEEs, data transfer across heterogeneous enclaves relies on the transit of non-secure regions, necessitating re-encryption and decryption to maintain data security. This re-encryption and decryption process introduces a substantial amount of extra memory access, which in turn leads to a competition between data transfer and computation for memory bandwidth, preventing parallel execution.

To address these issues, our work creates \textbf{a unified TEE} to efficiently support secure heterogeneous collaborative computing, eliminating VN overhead of intensive batching data accesses in CPU, computation stalls caused by large-granularity MAC in NPU, and re-encryption overhead of secure communication channel. 
1) For CPU, ~\name~ virtually supports the on-chip tensor-granularity VN management (compatible with cacheline granularity), thus eliminating the off-chip VN access and Merkle Tree traversal overhead. ~\name~ efficiently detects and manages on-chip tensor structures by leveraging the dimensional-wise streaming access pattern and flexible entries merging operation.
2) For NPU, ~\name~ propose tensor-wise MAC management with delayed verification to reduce storage overhead and eliminate computation stalls. It maintains data integrity with the proposed \textit{tensor poison tracing and verification barrier} mechanism to ensure that data tampering is limited in the NPU enclave and can soon be detected before communication. To ensure code integrity, we restrict instruction memory requests following the normal non-delayed verification dataflow.
3) For heterogeneous communication, the unified granularity of ~\name~ brings the compatibility of ciphertext across heterogeneous enclaves, avoiding cumbersome re-encryption and scheduling in data transfer.

The main contributions of the paper are as follows:
\begin{itemize}
    \item We analyze the overhead stems from different granularities between CPU and NPU in existing heterogeneous TEE designs, 
    which inspires our unified tensor granularity heterogeneous TEE designs.
    \item We propose~\name, a unified granularity heterogeneous TEE architecture, including the hardware-based tensor-granularity CPU TEE, the tensor-wise MAC management with delayed verification, and a direct data transfer protocol between secure enclave memory without re-encryption.
    \item We evaluated~\name~with enhanced Gem5 and a cycle-accurate NPU simulator. Our evaluation results show that~\name~improves the performance of LLM training by 4.0x compared to existing work and incurs only 2.1\% overhead compared to non-secure training.
\end{itemize}

\section{Background}
\label{sec:background}

\subsection{Collaborative Computing}
Due to the distinct computational model and limitations of CPUs and NPUs, collaborative computing is becoming increasingly popular~\cite{AlphaFold2021,ren2021zero, tesla}.

\begin{figure}
    \centering
    \includegraphics[width=0.95\linewidth]{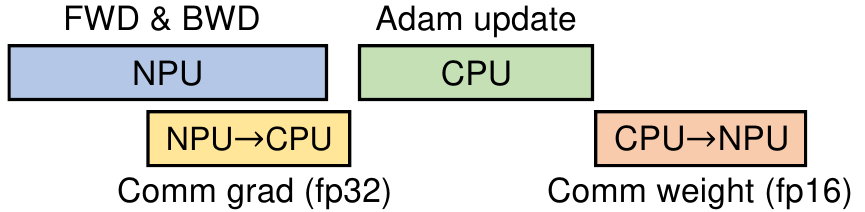}
    \caption{Zero-Offload dataflow. Light computation like weight update and high precision data like weights and optimization states are offloaded to CPU.}
    \label{fig:zero_offload_flow}
\end{figure}

\textbf{LLM CPU-NPU collaborative computing}: 
Large language model training has tended to heterogeneous collaborative computing with NPU and CPU due to the limited NPU DRAM capacity~\cite{ren2021zero,choukse2020buddy,hildebrand2020autotm}. 
This paper focuses on the ZeRO-Offload~\cite{ren2021zero} for evaluation study, which is a prevalent framework in collaborative training for LLM.
The main reason is that ZeRO-Offload could offload phases with lower computational loads but high memory usage from the NPU to the CPU which effectively mitigates memory storage pressure on the NPU, thus enabling larger model training. 
For example, by offloading the gradients and optimizer states to the CPU while keeping forward and backward computation on NPU, ZeRO-Offload enables a 10x increase in training model size.
As illustrated in Figure~\ref{fig:zero_offload_flow}, the computation flow of ZeRO-Offload mainly involves three stages.
First, the NPU performs forward and backward computations, passing the updated gradients to the CPU during backpropagation. 
Then, the CPU conducts optimizer iterations to update optimizer states and weights. 
Finally, the CPU transfers the updated weights back to the NPU. 

\subsection{Memory Protection}
In TEE, memory protection encompasses ensuring data confidentiality through encryption, integrity via Message Authentication Code (MAC) verification, and freshness with the use of a monotonic counter.

\textbf{Memory encryption}:
TEE typically employs counter mode AES for memory encryption, which introduces a counter that consists of a physical address (PA) and a version number (VN). The VN increases with each write-back, against replay attack.
Let $K_{AES},~P$ and C denote the AES encryption key, plaintext, and ciphertext respectively. 
Then, the encryption can be formulated as $C = AES(K_{AES}, (PA, VN)) \oplus P $, where $\oplus$ represents XOR.
The decryption follows the same procedure due to the XOR nature.

\begin{figure}
    \centering
    \includegraphics[width=1\linewidth]{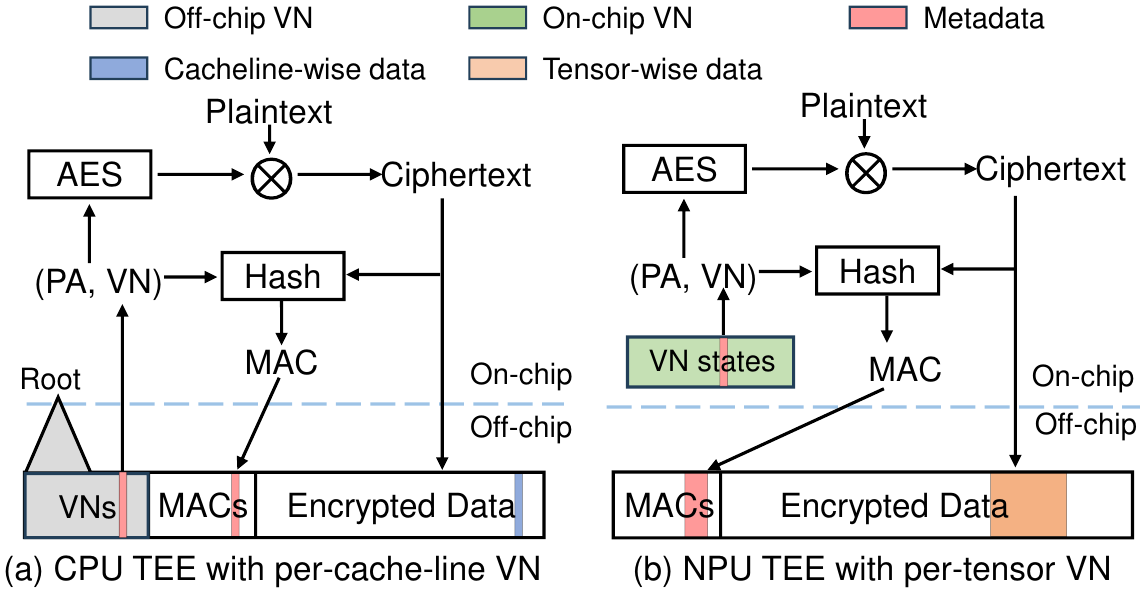}
    \caption{Two types of memory protection schemes. Tensor-wise protection eliminates the off-chip memory access for VN since the per-tensor VNs could be stored on-chip.}
    \label{fig:memory_protection}
\end{figure}

\textit{Mismatched granularity}: 
Figure~\ref{fig:memory_protection} illustrates two distinct granularities for VN management.
Firstly, the CPU holds a dedicated VN for each cacheline to support random memory access (Figure~\ref{fig:memory_protection} (a)). 
However, the fine granularity, such as a 56-bit VN for a 64B data block, introduces large storage overhead (11\%) and memory access overhead~\cite{costan2016sgx}. 
Moreover, off-chip VNs rely on the Merkle Tree to ensure their integrity. 
Secondly, recent works~\cite{hua2022mgx,shrivastava2023securator} embrace tensor-wise VN for NPU, leveraging the regular memory access patterns prevalent in NPU workloads (Figure~\ref{fig:memory_protection} (b)).
With tensor granularity, a single VN is used for all cachelines within the tensor.
This approach significantly reduces storage pressure and enables on-chip storage of VNs, thereby reducing memory access for VNs and eliminating the need for Merkle Tree.

\textit{Limitations of existing work}: SoftVN~\cite{umar2022softvn} preserves tensor structure within the on-chip VN Table by explicitly declaring it in the software code. It performs efficiently in simple static scenarios with minimal hardware overhead. However, SoftVN has certain limitations:
1) \textit{Inapplicable in complex and dynamic scenarios}: SoftVN~\cite{umar2022softvn} assumes a static and simple data flow in programs. 
However, in LLM training, various runtime parallel strategies and further complex operations like tensor splitting and merging~\cite{rajbhandari2020zero} exist, making it challenging to trace the data flow of tensors and specify VN at each writeback in the top-level user code.
2) \textit{Dilemma for improving practicability:} Since the acquisition of VNs occurs on critical paths (cache access) in SoftVN, increasing the manageable number of tensor entries results in performance degradation. This puts SoftVN in a dilemma between practicality and high performance under complex applications.
3) \textit{Wastage of Entries}: 
In scenarios where a single tensor is used in parallel across multiple cores, SoftVN leads to each subtensor occupying a VN Table entry in the corresponding core. 
This contributes to the exacerbation of capacity pressure, underscoring a critical limitation of the SoftVN.

\textbf{Integrity verification}:
To ensure data integrity, the MAC of the data is calculated and stored in DRAM during write-back. In subsequent reads the MAC is recalculated and validated to ensure that it is identical to the one from off-chip storage. The MAC calculation can be represented as $\text{MAC} = Hash(K_{MAC}, (C, PA, VN))$. To prevent replay attacks, the MAC should be stored on-chip. However, due to limited on-chip storage, VN and Merkle Tree are employed to reduce on-chip storage overhead. BMT~\cite{rogers2007BMT} further proposes that the Merkle Tree only needs to protect the VN, reducing the width of the Merkle Tree. Eventually, the root node of the Merkle Tree is securely stored on-chip while VNs and MACs are stored off-chip. The integrity verification for each memory access relies on recursive layer-by-layer verification of the Merkle Tree, resulting in significant overhead. For the NPU, the VNs are securely stored on-chip, eliminating the need for iterative access to the Merkle Tree.

\subsection{Heterogeneous NPU TEE}
Recent works have extended CPU TEE to heterogeneous TEE for mission-critical computations within NPU. Heterogeneous NPU TEE can be classified into integrated and discrete types based on their connectivity with the CPU.

\textit{Integrated NPU TEE}: Integrated architecture places both the CPU and NPU on the same chip and shares the same DRAM, thus with no problem for data transfer between CPU and NPU~\cite{lee2022tnpu,deng2022strongbox}. 
However, The integrated NPU TEE's performance is limited due to power and area constraints, which makes it insufficient for demanding applications such as LLM training that involve large computations.

\textit{Discrete NPU TEE}: Discrete heterogeneous architecture has separate CPU and NPU chips and connects them with data buses like GDDR. The discrete NPU, featuring dedicated controllers and memory, provides enhanced computational power to support LLM training. 
Existing works on NPU memory protection prefer tensor-wise VN management for performance: 
Common Counter~\cite{na2021common} scans memory during kernel switches, saving metadata on the chip for data regions with the same VN, which reduces additional memory access overhead for metadata. MGX~\cite{hua2022mgx} and Securator~\cite{shrivastava2023securator} on the other hand generate corresponding VNs for each memory access based on on-chip execution status during kernel execution, minimizing VN storage and access overhead. Since the discrete heterogeneous architecture is more widely adopted in the product environment, we will focus on the discrete heterogeneous TEE.


\subsection{Threat Model}
As a heterogeneous system, our Trusted Computing Base (TCB) only consists of the on-chip architectures (like CPU, NPU, and caches).   
The off-chip host memory and NPU GDDR memory reside outside the TCB. For NPU, following the same configuration with recent works~\cite{na2021common, hua2022mgx,abdullah2023plutus}, we consider NPUs with off-chip connected GDDR memory. The off-chip memory is vulnerable to physical attacks~\cite{yuan2022adaptive} like bus snooping attack~\cite{huang2002keeping,huang2014hmtt, CCA, shi2005towards} with mature toolkits already developed~\cite{hmtt_project} and cold boot attack as assumed by Intel TDX~\cite{TDX}, AMD SEV~\cite{kaplan2016amd}.

We assume the adversary has full control over the OS or privileged software and can carry out physical attacks. The adversary can snoop the bus (memory bus or PCIe bus)~\cite{huang2002keeping,huang2014hmtt} or employ cold boot attacks~\cite{lee2022tnpu, amdsev, TDX} to bypass security isolation for data theft. They can also manipulate the bus signals for data corruption or relay attacks.
We do not consider other side-channel attacks such as timing-based, power-based, and electromagnetic-based attacks~\cite{hunt2020telekine, karimi2018timing,lee2020off, naghibijouybari2018rendered, kadam2018rcoal}, as well as denial-of-service attacks and adversarial attacks~\cite{chakraborty2018adversarial,han2021scalecert,han2023real}. Model extraction attacks can be mitigated by orthogonal approaches~\cite{stefanov2018pathoram}.


\section{Motivation}
\label{sec:motivation}

\subsection{Inefficient CPU TEE Computation with Cacheline-Wise Metadata}

\begin{figure}
    \centering
    \includegraphics[width=1\linewidth]{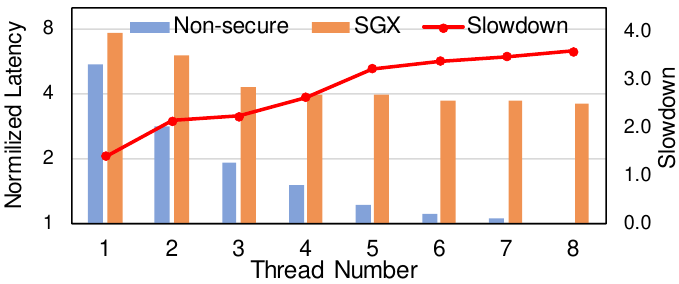}
    \caption{CPU TEE incurs large performance overhead and leads to the transition of the Adam workload from computation-intensive to memory-intensive due to additional memory access for metadata.}
    \label{fig:motivation_sgx_cost}
\end{figure}

In heterogeneous collaborative LLM training, the CPU is responsible for optimizer updates like the Adam optimizer step.
Figure~\ref{fig:motivation_sgx_cost} shows that up to 3.7x slowdown is caused when TEE is introduced. In the SGX environment, as the number of threads increases, the performance benefits gradually diminish, indicating a transition from compute-intensive to memory-intensive tasks. The increased memory access is mainly attributed to metadata such as VN and MAC.

\begin{figure}
    \centering
    \includegraphics[width=1\linewidth]{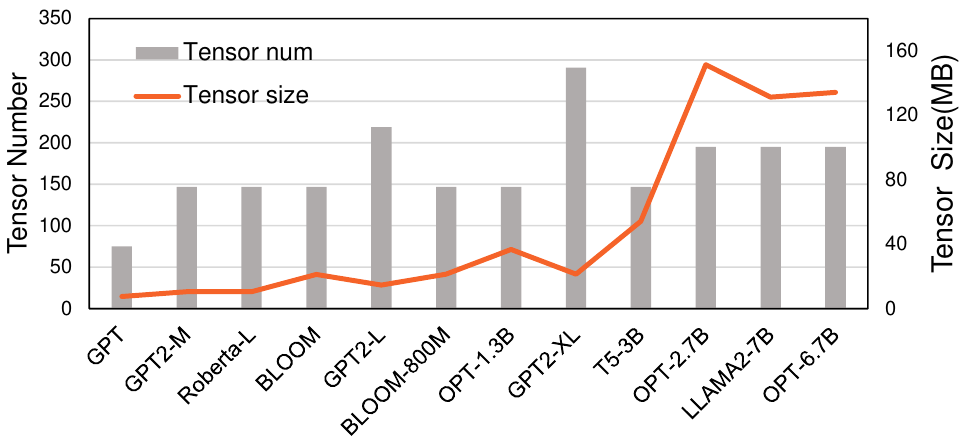}
    \caption{Access characteristics. The data is stored and accessed in tensor format, with small numbers and large sizes.}
    \label{fig:cpu_characteristic}
\end{figure}

For the Adam update, the tensor data are element-wisely accessed. The regular access pattern enables memory protection at the tensor granularity. 
Figure~\ref{fig:cpu_characteristic} illustrates the number and size of tensors during optimizer updates for different models. We observe that the tensor sizes grow to MBytes, but the growth rate of tensor numbers is slow, reaching only a few hundred. The data characteristic of small numbers and large sizes of tensors means that the memory protection of tensor granularity will achieve great performance gain with a small overhead.
However, CPUs always access memory at the granularity of cachelines, and the memory organization or instructions do not have tensor types. 

\subsection{Cacheline-Wise MAC of NPU Leads to Large Storage and Performance Overhead}
The memory resources on the NPU are limited, compared to CPU memory of up to 512GB on server CPUs, NPU memory capacity is usually only 40GB. The storage of enclave metadata further exacerbates the shortage of memory resources.
Recent NPU works with tensor-granularity VN~\cite{hua2022mgx,southsecure} reduce VN's storage overhead, yet close to 10\% of the storage overhead and 12\% performance overhead from MAC remains unresolved. Plutus~\cite{abdullah2023plutus} employs predictive execution to mitigate additional memory accesses for MAC, but it does not address the storage overhead. MGX~\cite{hua2022mgx} and GuardNN~\cite{hua2022guardnn} elevate the granularity of MAC from 64B to 512B, but they do not consider the computation stall issue resulting from the increased granularity (see Section~\ref{sec:delayed_verify}). Therefore, an integrity verification method that can reduce both storage and performance overhead is urgently needed.

\begin{figure}
    \centering
    \includegraphics[width=1\linewidth]{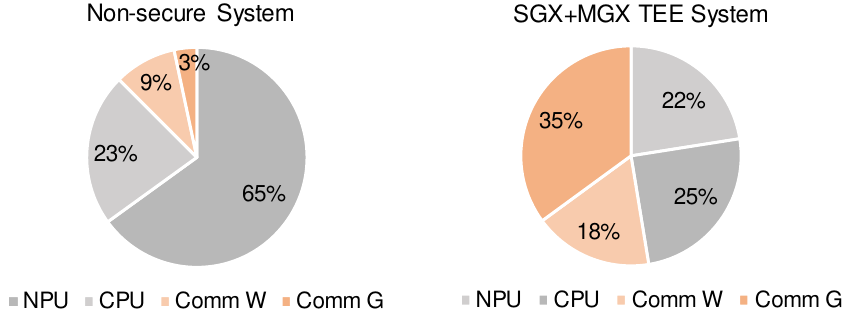}
    \caption{Breakdown of collaborative computing that GPT2-M model training with ZeRO-Offload. The communication occurs unacceptable overhead with TEE.}
    \label{fig:motivation_breakdown}
\end{figure}

To address the above challenges, we propose leveraging tensor-wise MAC with delayed verification and offer the following assurances for integrity with delayed verification: 1) Tampering can only occur on data but not code and tampered data can be soon detected. 2) Tampered data on the NPU cannot leave the NPU enclave, including the NPU chip and GDDR memory. 
Firstly, normal non-delayed verification of NPU code is used to prevent code tampering and render delayed-verification-based attacks~\cite{shi2005towards, shi2006authentication}.
Secondly, we ensure the integrity of communication data, preventing tampered data from leaving the NPU enclave.  Thirdly, though NPU data tampering may occur within a short time window due to delayed verification, it cannot be exploited to carry out more powerful attacks (e.g. data theft) since the NPU as a pure accelerator lacks file I/O abilities.

\subsection{Inefficient Data Transfer Protocols}
Figure~\ref{fig:motivation_breakdown} shows that the communication 
only occupies 12\% overhead in non-secure mode, but 
significantly increases to 53\% in the baseline system where CPU with SGX-like TEE and NPU with MGX-like TEE. The reasons are as follows:

\begin{figure}
    \centering
    \includegraphics[width=1\linewidth]{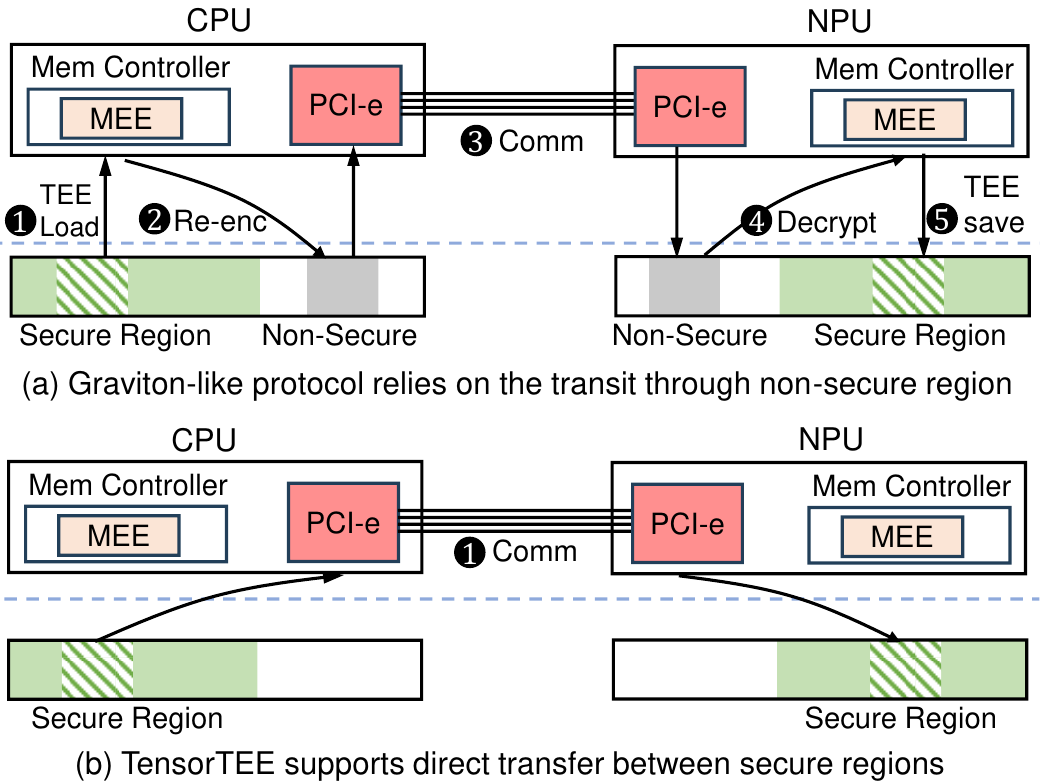}
    \caption{Comparison of data transfer protocols. ~\name~enables direct communication between secure memory.}
    \label{fig:motivation_protocol_comparison}
\end{figure}

Mismatched protection granularity of CPU and NPU incurs extra re-encryption. CPU and NPU have a mismatched granularity in that CPU performs protection on cacheline granularity while NPU on tensor granularity. Due to the granularity incompatibility, the encrypted data of CPU (or NPU) TEE can not be correctly decrypted by NPU (or CPU) TEE even with the same secret key since their fixed hardware engine and different metadata format. Therefore, communication relies on non-secure memory as a relay (Figure~\ref{fig:motivation_protocol_comparison} (a)). The sender needs to decrypt the data of the secure memory region and save it in the non-secure area. After communication, the receiver loads the data into the secure region. To protect the privacy of the data, the data in the non-secure region need to be re-encrypted. 

\begin{figure}
    \centering
    \includegraphics[width=1\linewidth]{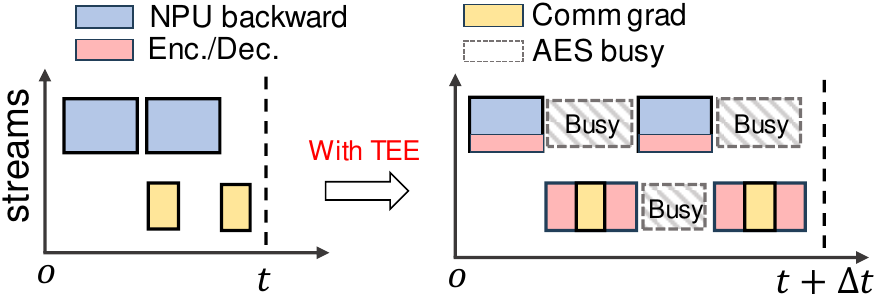}
    \caption{Bounded AES bandwidth hinders the parallel of computation and data transferring. The two rows represent the computation (top) and communication stream (bottom).} 
    \label{fig:motivation_serial_exe}
\end{figure}

The bandwidth contention caused by communication (re-encryption) and computation (IO read/write) results in both tasks being forced to execute sequentially (Figure~\ref{fig:motivation_serial_exe}), even when the encryption engines are sufficient. 
For resource-constrained NPUs, the hardware cost limits the deployment of multiple encryption engines, resulting in the contention of encryption bandwidth~\cite{zuo2021sealing}. Recent work SecureLoop~\cite{lee2023secureloop} reports that a fully-pipelined cryptographic engine accounts for nearly 35\% of the logic gates in an Eyeriss-like NPU. In our simulation, one AES engine cannot provide enough bandwidth even for NPU computing (8 GB/s provided, but at least 20 GB/s is needed), thus communication data have to be delayed and serialized with NPU computing.
Even for large NPUs with sufficient AES engines, re-encryption during communication exacerbates the pressure of memory bandwidth and blocks the IO relevant to computation. The re-encryption process is extremely memory-intensive and occupies the most DRAM bandwidth (84\%) in our simulation, hindering its parallelization with NPU computation.
Both of the above indicate inefficient dataflow caused by the sequential execution of computation and communication. In this paper, we assume each channel has its dedicated encryption engine to balance the trade-off between hardware overhead and bandwidth requirements.

\subsection{Optimization Goal}
These observations inspire a unified tensor-granularity heterogeneous TEE architecture for several benefits:
1) The tensor-wise CPU TEE with on-chip VN reduces memory access for metadata.
2) Tensor-wise MAC management of NPU mitigates storage and performance overhead while ensuring NPU data security.
3) The unified heterogeneous tensor-wise TEE enables direct data transfer between secure memory.

\section{\name~Design}
\label{sec:arch}

\subsection{Overview}

\begin{figure}
    \centering
    \includegraphics[width=1\linewidth]{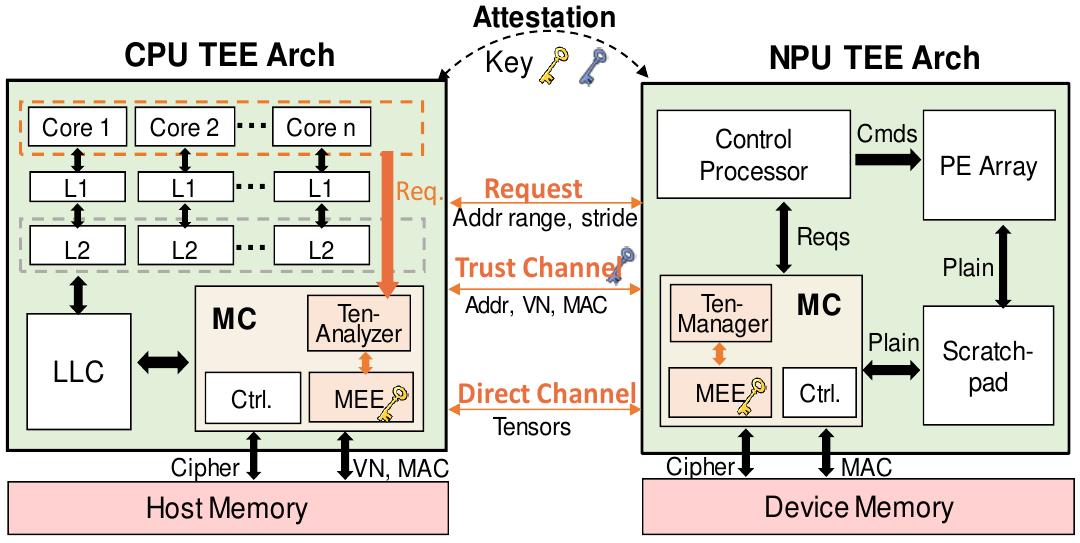}
    \caption{Architecture Overview and TCB of ~\name.}
    \label{fig:overview_arch}
\end{figure}

Our overall architecture, as shown in Figure~\ref{fig:overview_arch}, introduces three optimizations: 

\textit{Tensor-wise memory protection on CPU}: To unify the protection granularity and improve the CPU TEE performance on collaborative computing, we propose the hardware-based Tensor-Analyzer (TenAnalyzer) component for tensor-wise memory protection. TenAnalyzer receives and analyzes memory access requests from the CPU cores, constructing and maintaining the Meta Table data structure in run-time. Each entry of the Meta Table holds shared metadata (like VN) for all cachelines within a tensor, eliminating the overhead of metadata memory access in subsequent memory requests. Our design is optional and complementary to the traditional protection method like SGX~\cite{costan2016sgx}, enabling or disabling tensor management through the Enable Tensor-wise Management Flag (EnTMF) without affecting the software programming. In non-tensor applications, tensor-wise management is disabled with negligible overhead.

\textit{Tensor-wise MAC with delayed verification on NPU}: 
To improve both the performance and storage efficiency, ~\name~ proposes tensor-wise MAC management with delayed verification for only NPU tensor data. Delayed verification enables parallelization between verification and computation, eliminating stalls caused by coarse-grained MAC verification. To ensure integrity, ~\name~ 1) restricts code access following the normal non-delayed verification dataflow to maintain code integrity and prevent any potential attacks and 2) ensures the communication data integrity when they leave the NPU enclave by proposed \textit{tensor poison tracing and verification barrier} mechanism. 

\textit{Implementation of direct data transfer protocol}: Leveraging a unified tensor granularity, we enable an efficient data transfer protocol that eliminates the transit through non-secure regions and enables the parallel between data transfer and computation. The protocol contains two channels: A trusted channel is used for the encrypted transfer of on-chip metadata and a direct channel is for the transfer of ciphertext tensors between heterogeneous secure enclaves.

\subsection{CPU TEE with Tensor Granularity}
\label{sec:tensor_cpu_tee}

To increase practicability and alleviate the burden on programmers, a hardware-based tensor-wise CPU memory management mechanism is required. It should incorporate tensor management capabilities while maintaining low overhead.

\begin{figure}
    \centering
    \includegraphics[width=0.9\linewidth]{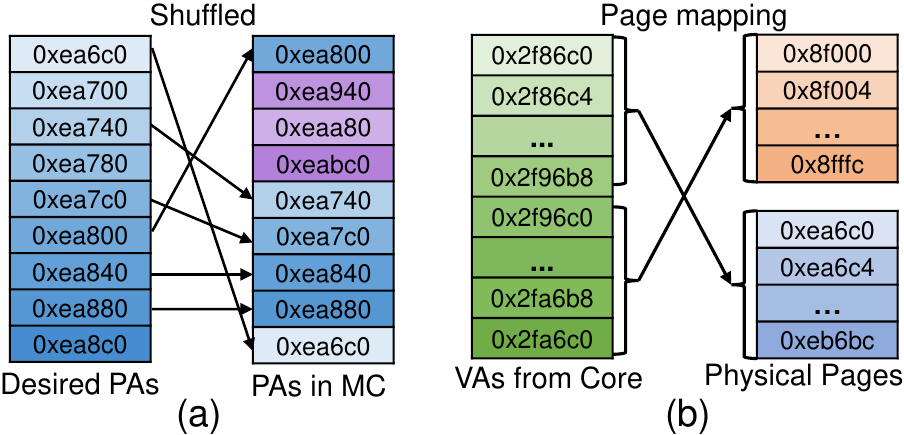}
    \caption{Memory access analysis. (a) The expected regular pattern on the physical address (PA) is shuffled by the cache. (b) Core virtual address (VA) requests remain regular and continuous covering all discontinuous physical pages.}
    \label{fig:trace_analysis}
\end{figure}

\textbf{Challenges of hardware-based tensor data structure detection}: 
To address the entry wastage issue, the tensor management structure should be located outside the cores, such as in the Memory Controller (MC). However, there are challenges with tensor detection and maintenance in the MC:
1) Cache interference: As shown in Figure~\ref{fig:trace_analysis}, the process switches lead to random cache line evictions, disrupting memory access patterns. Additionally, cache prefetchers like stride prefetchers shuffle the access order when the wrong prefetch request is issued.
2) MC receives physical addresses, and their contiguity is limited by the page size.
3) Long latency: The access of the large-capacity cache before the request reaches the MC introduces significant memory access latency, resulting in long latency for maintaining and accessing tensor data structures.

\textbf{Tensor structure management in CPU TEE}: 
To address the aforementioned challenges, we propose TenAnalyzer, located within the MC, which directly receives memory access requests from the cores, as shown in Figure~\ref{fig:overview_arch}. TenAnalyzer offers the following advantages:
1) It receives memory access requests directly from the cores, avoiding cache interference from evicting and prefetching as shown in Figure~\ref{fig:trace_analysis} (a).
2) Leveraging virtual addresses from the cores maximizes the utilization of tensor contiguity as shown in Figure~\ref{fig:trace_analysis} (b). To avoid the same VA problems, each enclave uses its dedicated key and the Meta Table is saved and restored for context-switching cases.
3) Detection, maintenance, and access of tensor data structures can be parallelized with multi-level caches, effectively hiding their access latency.
Thus, TenAnalyzer overcomes the limitations of SoftVN~\cite{umar2022softvn} and provides an efficient and scalable solution for tensor-based memory management. 

\begin{figure}
    \centering
    \includegraphics[width=0.9\linewidth]{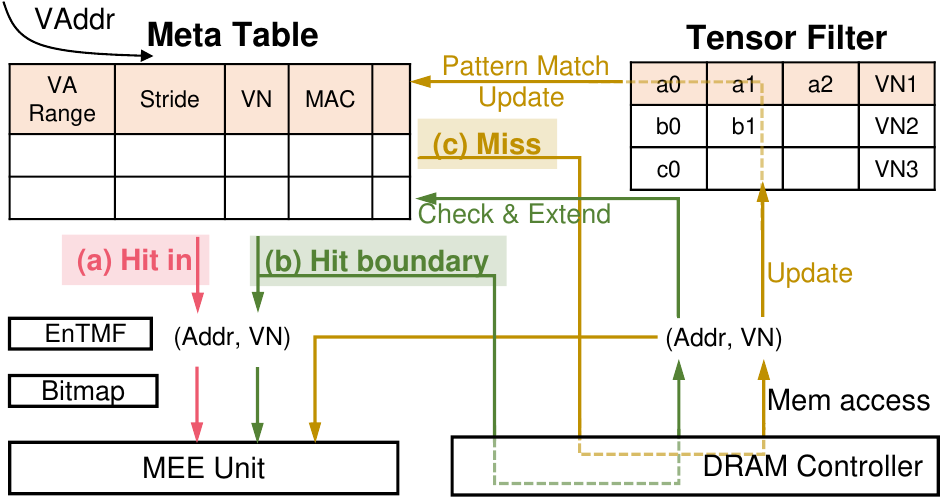}
    \caption{Reading dataflow for tensor detection. Tensor Filter collects address set and initializes entry in Meta Table. Meta Table receives requests and provides metadata to the MEE Unit.}
    \label{fig:reading_dataflow}
\end{figure}

TenAnalyzer mainly contains two components: Meta Table and Tensor Filter, as shown in Figure~\ref{fig:reading_dataflow}. 
\underline{1) Meta Table} maintains the tensor structure and can be constantly updated by boundary hit. Specifically, the Meta Table receives the access request of the core, and when it hits, the address and VN are passed to the MEE for encryption, which avoids the extra access for metadata in memory. 
\underline{2) Tensor Filter} handles the Meta Table misses, collecting and analyzing the request addresses. When a specific entry in the Tensor Filter reaches its collection address limit (4 in our setting), it checks if the entry meets the tensor condition: having the same VN and a consistent pattern between addresses. The successfully checked entry is then populated into the Metadata Table as the initialization structure for tensors. \underline{Besides}, other components, a bitmap and two flags for indicating the cacheline update states and the EnTMF flag for enabling or disabling TenAnalyzer, are used.

\textbf{Dataflow for reading (or detection)}: As shown in Figure~\ref{fig:reading_dataflow}, TenAnalyzer has three cases when a request arrives: \textit{hit in}, \textit{hit boundary}, and \textit{miss}. \underline{\textit{1) Hit in}} means that the request hits the address range of an entry in the Meta Table, and then the corresponding VN will be transmitted to MEE to start the encryption/decryption. \underline{\textit{2) Hit boundary}} means that the request hits the boundary of the address range of an entry (request address == last address + stride) in the Meta Table that the VN cannot guarantee to be correct. At this time, assuming that this VN is correct, the VN is still sent to MEE, and at the same time, the VN access in DRAM is started. If the VN in DRAM is consistent with the assumed VN, \textit{the relevant Meta Table entry is updated and its address range is expanded, which leads to gradual coverage of tensor detection}. \underline{\textit{3) Miss}} means that there is no matching in the Meta Table, then the VN will be obtained by the DRAM access and this request is sent to the Tensor Filter for subsequent pattern detection and filtering. For reading, the on-chip tensor-wise VN always remains equivalent to the off-chip cacheline-wise VN while efficiently providing VN to minimize off-chip memory access and Merkle Tree traversal.

\begin{figure}
    \centering
    \includegraphics[width=1\linewidth]{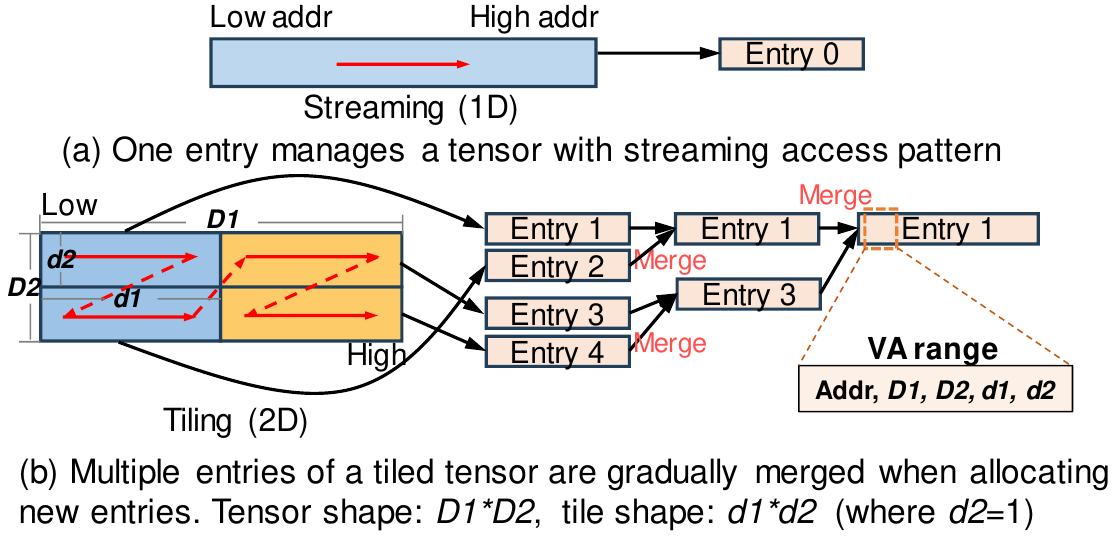}
    \caption{TenAnalyzer enables detecting tensors with complex access patterns based on dimensional-wise streaming access
pattern and flexible entries merging.}
    \label{fig:entry_merge}
\end{figure}

TenAnalyzer efficiently detects tensors with complex access patterns through entries merging. For streaming operations, such as Adam's update, a single entry is sufficient for the detection and management of large tensors (Figure~\ref{fig:entry_merge} (a)). However, for tensor operations requiring tiling, such as matrix multiplication, detecting complete tensor information with a single entry becomes challenging. Nevertheless, TenAnalyzer can still manage each short streaming data like a row of a tile and infer the tile dim (e.g. the first blue row with a size of \textit{d1}) based on the dimensional-wise streaming access pattern. 
Next, TenAnalyzer attempts to merge a few recently updated entries when creating new entries. The merge operation is not restricted to address-adjacent tiles but allows merging in multiple directions, enabling more timely merges and more efficient on-chip storage (2 directions for 1D tensor, 4 directions for 2D tensor and 6 directions for 3D tensor). To merge tiles with non-adjacent addresses, it's required that the tile dims, stride, and VN match. The successful merge allows inferring the tensor dimension (e.g. merging the two blue rows infers the tensor dimension \textit{D1}). The inferred dimension can serve as a constraint for subsequent merges to improve the merge accuracy.
By iterating this process, unified management of the tiled tensors can be achieved as shown in Figure~\ref{fig:entry_merge} (b).

TenAnalyzer gradually completes the tensor structure but with extra memory access to verify its correctness. 
Since the data transfer instructions from NPU typically include tensor structure information, such as address, size, and stride, \name~utilizes them to update the Meta Table entry with a large overlapped address range to speed up the tensor structure creation in the CPU.

\begin{figure}
    \centering
    \includegraphics[width=0.9\linewidth]{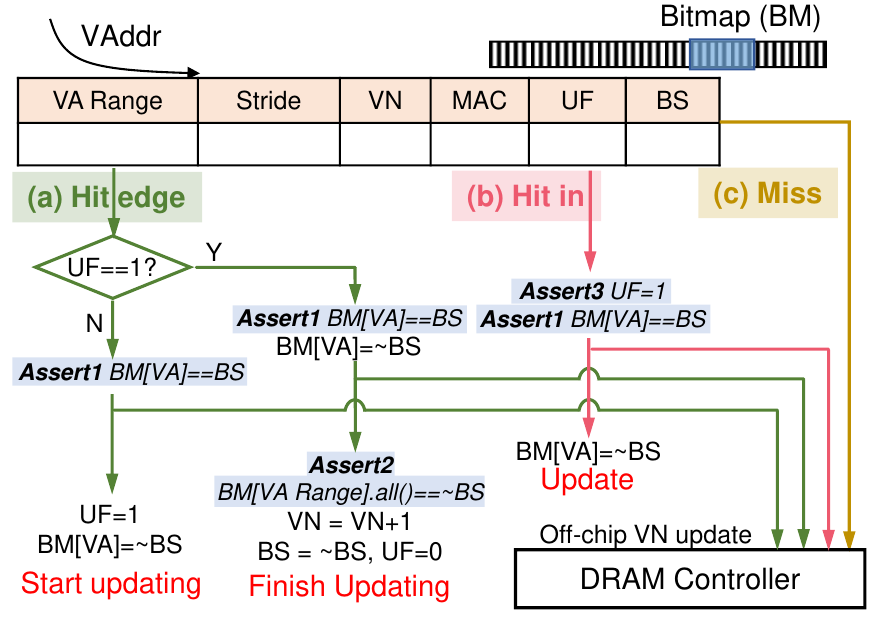}
    \caption{Writing dataflow of TenAnalyzer ensures the correctness of on-chip tensor VN updates and maintains consistency with off-chip cacheline-granularity VN.}
    \label{fig:writing_dataflow}
\end{figure}

\textbf{Dataflow for writing (or update)}: During tensor writing, TenAnalyzer efficiently maintains the correct updates of tensor VN by confirming each cacheline within a tensor must update only once within a tensor update using a bitmap (BM), Updating Flag (UF) and Bit State (BS). 
The bitmap and address range log cacheline updates for a tensor, flipping bits at addresses upon write requests. UF denotes an updating tensor, and BS shows the pre-update bitmap bit value. TenAnalyzer increments the VN after updating all tensor cachelines (all relevant bitmap bits flip). Hardware isolation mechanism secures the full bitmap in DRAM, and a small on-chip cache enhances bitmap access efficiency~\cite{feng2023mmt}.

TenAnalyzer encompasses 3 scenarios in tensor writing as illustrated in Figure~\ref{fig:writing_dataflow}: \textit{Hit edge, Hit in and Miss}. \underline{\textit{1) Hit edge}} means the request hits the first address (start updating) or last address (finish updating) which matches the common tensor-based application and allows complex memory access like tiling. The bitmap bit flips after checking with BS to be equal, marking the cacheline updated.
When the last cacheline address arrives, all bitmap bits in the address range are checked to be equal to the flipped BS to indicate the tensor updating finish; then the on-chip tensor increments, BS flips, and UF resets.
\underline{\textit{2) Hit in}} means the request hits the address range but not the edge. After checking the UF and bitmap state, we flip the bitmap bit.
\underline{\textit{3) Miss}} indicates out-of-range addresses, requiring only off-chip VN updates. Note that the writing addresses from cores are filtered by LLC to match the off-chip data update. Writing is performed on background, and not on the critical path.

The assertions (\textit{Assert1, 2 and 3}) ensure that each cacheline must be updated only once before the tensor update finishes to guarantee the correct update of on-chip tensor-wise VN. Assert1 ensures that each cacheline can not be updated before the tensor update starts, while Assert2 ensures that each cacheline must be updated once when the tensor update finishes. The access pattern is common in tensor-based applications.
When the assertion is violated, the Meta Table entry is invalidated, preventing inconsistencies between the tensor VN and the off-chip VN in the following corner cases: 1) The detected tensor is mixed with non-tensor variables. 2) An entry contains multiple tensors with varying update frequencies.

\subsection{Delayed Verification for Tensor-Wise MAC}
\label{sec:delayed_verify}
\begin{figure}
    \centering
    \includegraphics[width=0.95\linewidth]{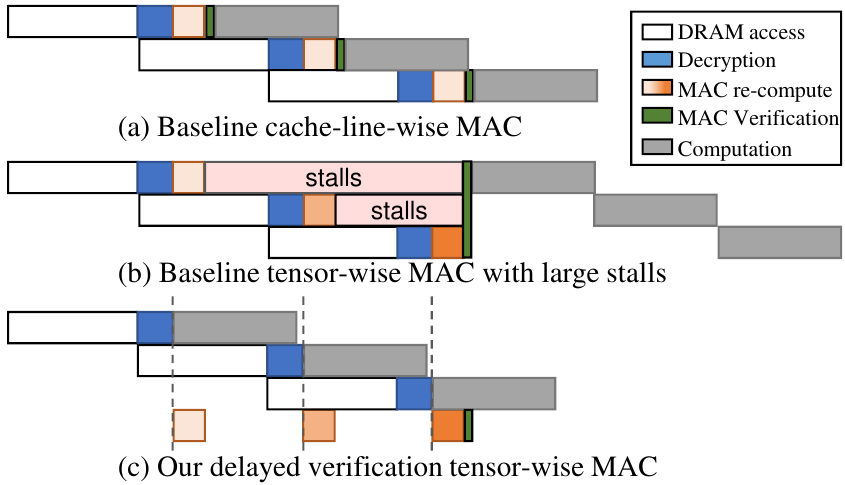}
    \caption{Pipeline comparison. (a) Baseline pipeline. (b) Later verification of tensor-wise MAC leads to large computing stalls. (c) Delayed tensor-wise verification avoids computation stalls.}
    \label{fig:delayed_verification}
\end{figure}

The granularity of MAC trades off the storage overhead and verification overhead. Cacheline-wise MAC results in large storage overhead in the NPU, which further hinders the deployment of LLM workloads on the NPU due to the limited memory capacity. 
Recent studies~\cite{hua2022mgx,hua2022guardnn}, have opted for a MAC granularity of 512 bytes, thus establishing a balance between storage overhead and verification cost. However, the use of large granularity MAC results in later verification, leading to computation stalls for already decrypted cachelines as shown in Figure~\ref{fig:delayed_verification} (b). 
Specifically, despite having the same time for MAC re-generation at both granularities, the blocked computation by later MAC verification stages results in many pipeline bubbles, causing the inefficient pipeline (13\% overhead with 4KB granularity).

To reduce storage overhead while avoiding computing stalls, leveraging the characteristics of LLM collaborative computing, we propose a tensor-wise MAC management method with delayed integrity verification. We parallel the kernel computation for decrypted data and re-generation tensor MAC as shown in Figure~\ref{fig:delayed_verification} (c). After completing the tensor-wise MAC, we perform the integrity verification for the entire tensor. In this way, we significantly reduce the storage overhead and memory traffic while avoiding large bubbles caused by the long stalls as shown in Figure~\ref{fig:delayed_verification} (b). For tensor-wise MAC calculations, we adopt a simple and efficient algorithm that involves XORing the MAC values of all cachelines which is similar to the MEE~\cite{gueron2016mee}. The algorithm can be formulated as follows where $MAC_{i}$ means the MAC for i-th cacheline in the tensor and $n$ is the number of cache lines:
$
    MAC_{tensor} = MAC_{0} \oplus MAC_{1} \oplus ... \oplus MAC_{n-1} $.
The XOR-based algorithm is insensitive to order, allowing various optimizations in NPU computing like tensor tiling.

\begin{figure}
    \centering
    \includegraphics[width=1\linewidth]{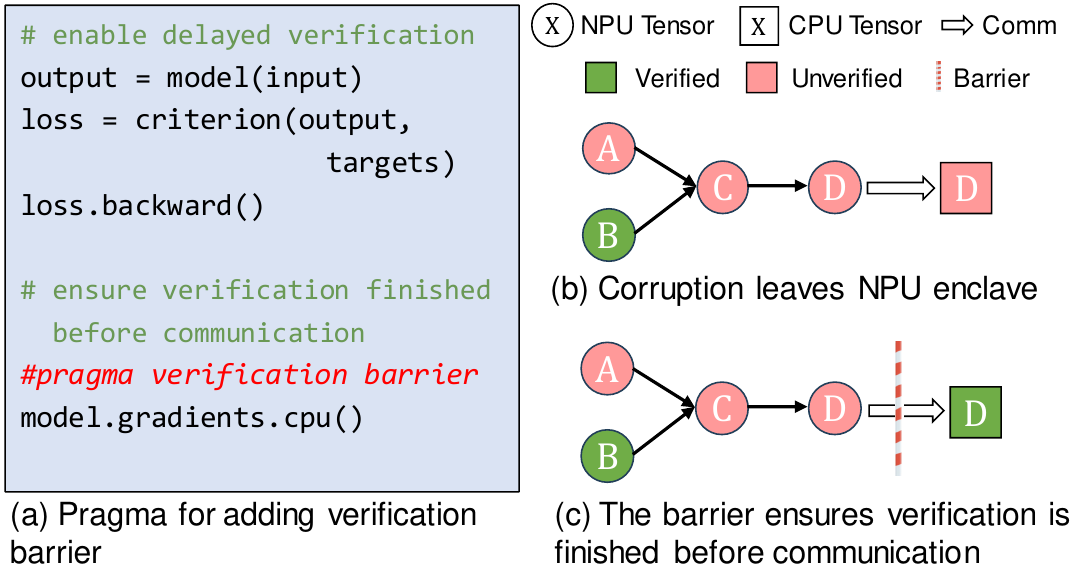}
    \caption{Protecting the data integrity in NPU communication with delayed verification by the verification barrier.}
    \label{fig:integrity_protection}
\end{figure}

\textbf{Integrity guarantee}: We ensure the integrity of tensor data on NPU computation with a strict integrity guarantee on code access and communication data. This allows for the delayed verification of intermediate tensors during computations, resulting in performance improvements without compromising security. Delayed verification allows temporary tampering on intermediate tensors, but can detect them by MAC soon. The temporary tampering of NPU intermediate tensors is acceptable as it cannot be exploited to carry out more powerful attacks, such as data theft. This is because the NPU, being a pure accelerator, lacks the privileged software (like OS) and file I/O abilities. In contrast, the NPU code and communication data that leave the NPU and GDDR memory need strict integrity guarantees.

\name~ guarantees code integrity by restricting code access requests following normal non-delayed MAC verification dataflow.
The non-delayed verification of NPU code disables attacks for delayed verification based on code tampering, such as pointer conversion, binary search, or disclosing kernel~\cite{shi2006authentication}.
We differentiate code and data requests in the memory controller and disable delay verification for code requests. The desired differentiation can be easily accomplished by adding an \textit{isInst} flag in the instruction request packet during issuance by the fetch component or last-level instruction cache. Specifically, in this paper, the indicator flag is provided by the instruction buffer (as assumed TPU-like architecture~\cite{jouppi2017tpu_v1}).

\name~ guarantees the integrity of NPU communication data by ensuring the completion of verification on the involved tensors before the communication. This is accomplished through two steps: tensor poison tracing and verification barrier. 
\underline{Firstly}, drawing inspiration from~\cite{lehman2016poisonivy}, we incorporate a poison bit for tensors to track poison propagation as shown in Figure~\ref{fig:integrity_protection} (c). The poison bit of unverified tensors is set to 1, and the potential poison effect propagates to the output tensors. The poison bit is cleared after the verification is finished.
\underline{Secondly}, we introduce the pragma \textit{verification\_barrier} inserted in the code before the communication, as depicted in Figure~\ref{fig:integrity_protection} (a). This pragma produces an integrity synchronization instruction at compilation that blocks the subsequent communication instructions until the poison bits of the involved tensors are cleared. To automate synchronization instruction insertion and reduce manual overhead, we identify specific functions such as $.cpu()$ at compilation.
Compared to Figure~\ref{fig:integrity_protection} (b), this tensor poison tracing + verification barrier mechanism successfully and efficiently prevents the corruption of communication data caused by tampered tensors.  Besides, the number of unverified tensors is limited with a counter to avoid meaningless computations after verification failure is detected.


\textbf{Security analysis:} Given that the XOR operation does not reduce the output space of MAC (56 bits), the forgery resistance at the tensor granularity does not have a significant decrease~\cite{gueron2016memory,wegman1981new}. Repeated "blind guessing" still requires nearly $2^{56}$ attempts to forge successfully.

\subsection{Direct Data Transfer Protocol Implementation}
\label{sec:comm_protocol}

In this section, we present the implementation of an efficient heterogeneous TEE data transfer protocol that eliminates the need for re-encryption and decryption and enables the parallel between computation and data transfer. 

\subsubsection{Optimizations Based on Unified Tensor Granularity} 


\begin{figure}
    \centering
    \includegraphics[width=1\linewidth]{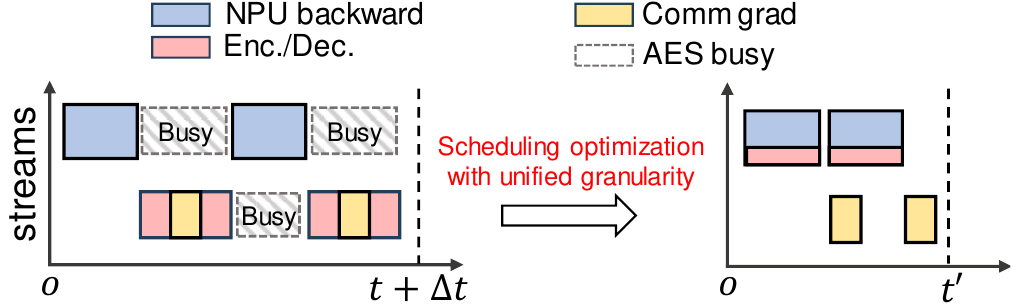}
    \caption{Unified-granularity design removes re-encryption and recovers parallel execution of computation and data transfer.}
    \label{fig:scheduling_optimized}
\end{figure}

Existing heterogeneous CPU and NPU TEE have different memory protection granularity, which leads to significant overheads in data transfer between the enclaves: First, the data transfer relies on a non-secure memory region relay, which incurs re-encryption overhead as shown in Figure~\ref{fig:motivation_protocol_comparison} (a). Second, since both computation and data transfer (with re-encryption) are memory-intensive, limited AES/memory bandwidth makes data transfer not parallel with computation as shown in Figure~\ref{fig:motivation_serial_exe}.

In this work, we reduce the data transfer overhead based on our proposed unified tensor management granularity. On the one hand, CPU and NPU enclaves have a unified tensor data structure, which means the off-chip encrypted memory could be decrypted on both sides. Thus, data transfer between the two does not need to rely on the transfer in non-secure regions, which makes it possible for direct data transfer between secure regions as shown in Figure~\ref{fig:motivation_protocol_comparison} (b). On the other hand, after eliminating the need for AES for data transfer, we break the AES/memory bandwidth bound which hinders parallel scheduling optimization. As shown in Figure~\ref{fig:scheduling_optimized}, based on the unified management granularity, we hide the data transfer in computation, greatly reducing the data transfer overhead.

\subsubsection{Implementation}
The protocol primarily consists of two phases: authentication and data transfer.

\textbf{Authentication phase}: The CPU and NPU build trust by remote attestation. Firstly, at the beginning of the execution of the workload, the CPU starts the process of enclave creation which copies code and data copy from non-secure memory to secure memory and computes the report of the enclave. Secondly, the CPU enclave sends the enclave creation request to NPU to create an NPU enclave with a similar process. Thirdly, the two enclaves authorize each other based on the report. 
To enable low-cost data transfer, the CPU and NPU TEE hold the same key for encryption/decryption. After the attestation, the two enclaves perform a key-exchange protocol like the Diffie-Hellman which enables the same key in both enclaves without leaking the key in the communication process. After the exchange, the key is always kept in the on-chip memory to ensure security.

\textbf{Communication phase}: There are two types of transfers: trusted metadata transfer and direct tensor transfer. We take the example of CPU-to-NPU data transfer to describe the process. As shown in Figure~\ref{fig:overview_arch}, in the metadata transfer phase, the NPU sends a data transfer request containing the tensor's address range. The CPU memory controller queries the Meta Table for metadata. The obtained tensor VN, MAC, and address are transmitted to the NPU through a trusted encrypted channel for subsequent decryption and verification. Simultaneously, the tensor data corresponding to the address range is directly transferred from the CPU DRAM to the NPU's DRAM via a direct channel, without involving the CPU and NPU. It is important to note that the two stages can be executed in parallel, requiring only the synchronization mechanisms upon completion of the transfers.
\section{Experimental Methodology}
\label{sec:exp_setting}

\subsection{Simulation Infrastructures}

We perform detailed simulations with the enhanced CPU simulator Gem5~\cite{binkert2011gem5}, an implemented cycle-accurate NPU simulator integrated with Ramulator~\cite{kim2015ramulator}, and the modeled system communication protocols with measured bandwidth. Further, we conduct performance alignment with SGX CPU and A100 GPU separately to ensure the correctness of the time models of our customized Gem5 and NPU simulator.

\begin{table}
    \centering
    \caption{System Simulation Configurations}
    \label{tab:hardware_config}
    \resizebox*{0.70\linewidth}{0.32\textheight}{
    \begin{tabular}{|l|l|}
    \hline \multicolumn{2}{|c|}{ \textbf{CPU Configuration} } \\
    \hline Frequence        & 3.5 GHz \\
    \hline Processors       & 8 out-of-order cores \\
    \hline L1 I/D cache     & 32KB, 8ways \\
    \hline L2 cache         & 256KB, 8ways \\
    \hline L3 cache         & 9MB, 8ways \\
    \hline DRAM             & DDR4@2400, 2 channels \\
    \hline Metadata cache     & 32KB  \\
    \hline AES Encryption   & 128-bit, 40 cycle lat. \\
    \hline MAC              & 40 cycle lat.  \\
    
    \hline \multicolumn{2}{|c|}{ \textbf{NPU Configuration} } \\
    \hline Frequence        &  1 GHz          \\
    \hline PE array         &  512*512        \\
    \hline Scratchpad       &  32MB           \\
    \hline DRAM             &  GDDR5, 40 GB, 128 GB/s \\
    \hline AES Encryption   &  40 cycles lat.  \\

    \hline \multicolumn{2}{|c|}{ \textbf{Commnicution Configuration} } \\
    \hline Comm. Bus & PCIe 4.0*16 \\

    \hline
    \end{tabular}
    }
\end{table}

\textbf{CPU Simulator}:
To evaluate the performance of CPU workloads on baseline SGX and~\name, we extend the Gem5~\cite{binkert2011gem5} simulator with the baseline SGX-like and our proposed tensor-wise memory protection mechanism. The CPU configurations are shown in Table~\ref{tab:hardware_config}. Gem5 is a cycle-accurate event-driven CPU performance simulator that supports multi-core, full-system, and memory simulation. We implement a scheme similar to Intel SGX MEE~\cite{gueron2016mee} as the baseline, which includes a 56-bit VN and MAC per 64-byte cacheline and an 8-ary Merkle tree. For SGX, each memory request arriving at memory control triggers additional accesses for VN and MAC. The MEE performs decryption for read requests and encryption for write requests. The MEE is augmented with a metadata cache to relieve the access pressure for accessing VN. For~\name, we implement and simulate the Meta Table and Tensor Filter extensions. To satisfy the requirement of collaborative computing, we select multi-core out-of-order CPUs with multi-channel large-capacity DRAM.

\textbf{NPU Simulator}:
To evaluate the performance of the NPU part computation, we implement a cycle-accurate simulator integrated with the DRAM simulator Ramulator~\cite{kim2015ramulator} and extend it with memory protection mechanisms. The NPU configurations are shown in Table~\ref{tab:hardware_config}. The implemented NPU simulator adopts the architecture of TPUv3 and employs an output stationary dataflow and incorporates automatic tiling and inter-layer optimization. Through simulation comparison, our simulator achieves computational performance similar to that of the A100 GPU. We extend the NPU simulator with MGX-like~\cite{hua2022mgx} on-chip VN management and memory protection component between NPU and DRAM.

\textbf{Communication}:
We model the basic Graviton-like~\cite{volos2018graviton} protocol and our direct data transfer protocol for performance evaluation and connect the communication simulations together with CPU and NPU simulators. We utilize the commonly used PCIe 4.0*16 as the communication bus between the CPU and NPU. 

\subsection{Workload}

\begin{table}
\centering
\caption{Workloads and Parameters}
\label{tab:workloads}
\resizebox*{0.6\linewidth}{0.22\textheight}{
\begin{tabular}{|c|c|c|}
\hline
Model      & \# Params & batch\_size \\ \hline
GPT        & 117M   & 60          \\ \hline
GPT2-M     & 345M   & 22          \\ \hline
Roberta-L  & 355M   & 22          \\ \hline
BLOOM      & 560M   & 21          \\ \hline
GPT2-L     & 774M   & 11          \\ \hline
BLOOM-800M & 800M   & 17          \\ \hline
OPT-1.3B   & 1.3B   & 10          \\ \hline
GPT2-XL    & 1.6B   & 6           \\ \hline
OPT-2.7B   & 2.8B   & 6           \\ \hline
XGLM-4.5B  & 4.5B   & 3           \\ \hline
LLAMA2-7B  & 6.7B   & 2           \\ \hline
OPT-6.7B   & 6.7B   & 2           \\ \hline
\end{tabular}
}
\end{table}

~\name~ is suitable for any tensor-based applications.
In this paper, we take one of the most representative and challenging workloads, LLM training, for the evaluation study. For LLM training, we select several popular models as our workloads, with parameter numbers ranging from 100M to 7B. The models and their configurations are presented in Table~\ref{tab:workloads}. We employ the scheduling strategy of the DeepSpeed framework with offload optimization for collaborative heterogeneous computing. In this paper, we focus on the LLM training scenario with a host multi-core CPU and a single NPU. To accommodate the memory capacity of NPU, we choose different batch sizes for different models. Besides, we choose 2D matrix multiplication with tiling optimization (GEMM) as CPU workload to show the effectiveness of ~\name~ on workloads with complex access patterns.

We compare three types of configurations in our experiments: 1) Non-Secure: In this setup, both the CPU and NPU do not utilize isolation and memory protection mechanisms like TEE. This configuration is used as a performance reference. 2) CPU SGX + NPU MGX configuration: In this setup, the CPU employs an SGX-like cacheline granularity protection mechanism, while the NPU utilizes an MGX-like tensor granularity protection mechanism. The communication between the CPU and NPU requires additional encryption and decryption due to granularity mismatch. 3) \name (ours): In this setup, we leverage the unified tensor-based management approach and optimization techniques proposed in our paper to achieve enhanced performance while maintaining security.

\section{Experiment Results}
\label{sec:exp_results}
In this section, we will first present the overall performance and performance breakdown of \name. Subsequently, we will delve into a detailed analysis of the benefits and effects of proposed optimizations.

\subsection{System Performance Improvement}

\begin{figure}
    \centering
    \includegraphics[width=1\linewidth]{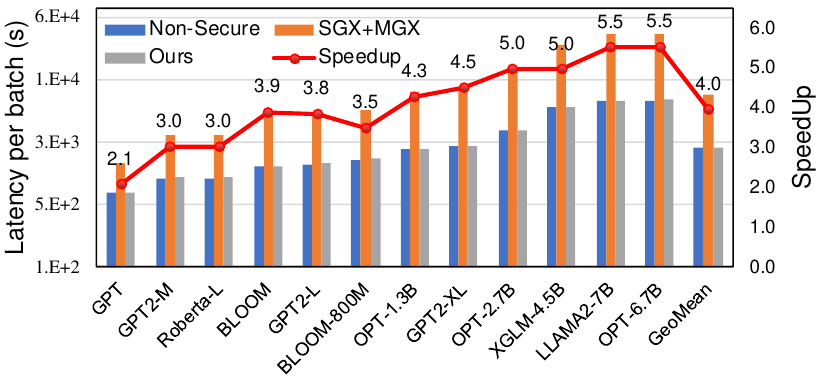}
    \caption{Overall performance comparison. ~\name~achieves significant performance improvement.}
    \label{fig:overall_performance}
\end{figure}

\begin{figure}
    \centering
    \includegraphics[width=1\linewidth]{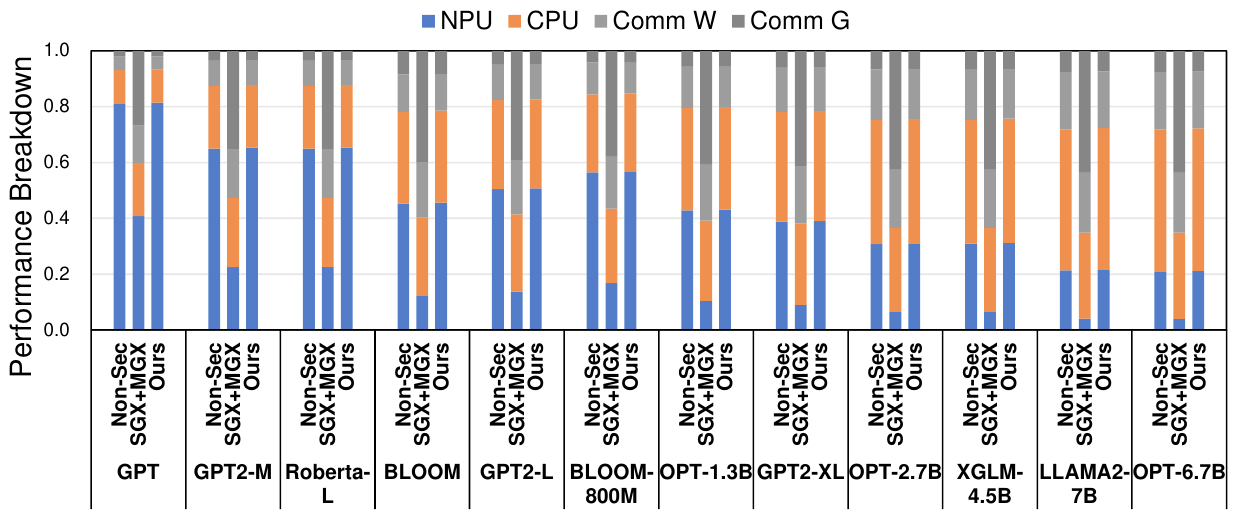}
    \caption{Bottleneck analysis. \name~effectively eliminates CPU computation overhead and data transfer overhead.}
    \label{fig:performance_breakdown}
\end{figure}

Figure~\ref{fig:overall_performance} showcases the overall performance comparison of the three methods across different model setups, while Figure~\ref{fig:performance_breakdown} provides a breakdown of the performance. The following observations can be made from these two figures:
1) Figure~\ref{fig:overall_performance} reveals that the baseline SGX+MGX method incurs a considerable overhead. Additionally, Figure~\ref{fig:performance_breakdown} depicts that this overhead primarily stems from the communication process of weights and gradients.
2) \name~ exhibits a significant performance improvement compared to the baseline. As seen in Figure~\ref{fig:overall_performance}, \name~ achieves a maximum performance boost of up to 5.5x and an average improvement of 4.0x compared to the baseline. Furthermore, when compared to the non-secure configuration, the average performance overhead of \name~ is only 2.1\%.
3) The Figure~\ref{fig:overall_performance} demonstrates a clear increasing trend of performance improvement for \name~ as the model size increases. The large models result in increased communication volume and CPU computation, as indicated by the breakdown in Figure~\ref{fig:performance_breakdown} which provides \name~ with greater optimization opportunities, leading to large performance gains.

\subsection{Improvement of Tensor-Wise CPU TEE}

\begin{figure}
    \centering
    \includegraphics[width=1\linewidth]{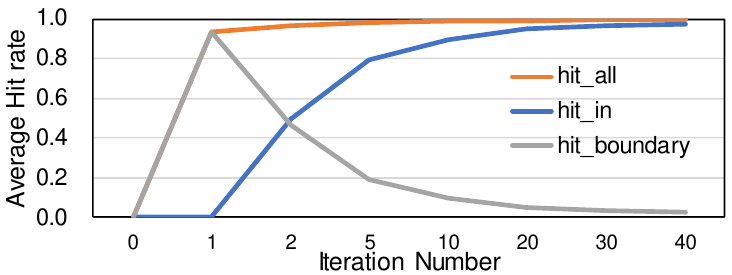}
    \caption{Meta table exhibits a high average hit rate within a few iterations.}
    \label{fig:vn_table_hit}
\end{figure}

The LLM training needs thousands of iterations to update the parameters. Figure \ref{fig:vn_table_hit} illustrates the two types of hits in the Meta table when the CPU performs the Adam optimizer after different iterations. 
\textit{hit\_all} represents either \textit{hit\_in} or \textit{hit\_boundary}. It can be observed that average \textit{hit\_all} is already sufficiently high after only one iteration (tensor detection is essentially completed) and gradually converges to 1 as the iteration count increases. As for average \textit{hit\_in}, it reaches 80\% by the 5th iteration and 95\% by the 20th iteration. The rapid convergence and high accuracy of Meta table access demonstrate the effectiveness of ~\name.

\begin{figure}
    \centering
    \includegraphics[width=1\linewidth]{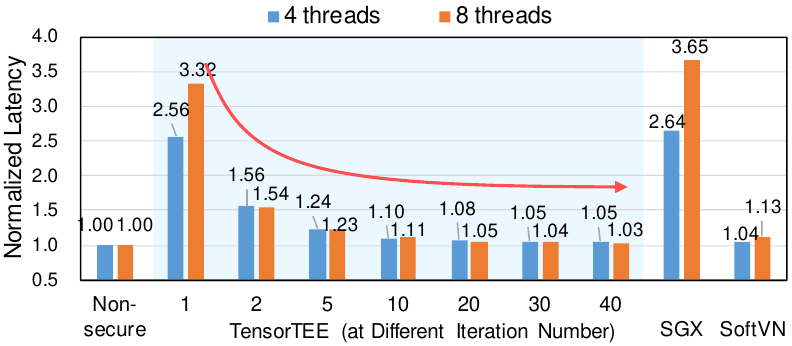}
    \caption{CPU performance comparison. TensorTEE achieves comparable performance with SoftVN in a few iterations.}
    \label{fig:cpu_performance}
\end{figure}

Figure \ref{fig:cpu_performance} presents the average performance overhead of \name's proposed VN tensor-based management compared to SGX and SoftVN at different iterations. For SoftVN, meticulous management of the VN table is required, including timely release and careful creation due to the limited number of entries. The following observations can be made:
1) The baseline SGX method incurs significant overhead, reaching a factor of 3.65x at 8 threads. The encryption and decryption overhead of SGX, along with additional memory accesses, significantly impact the memory-intensive Adam workload on the CPU. Furthermore, as the number of threads increases, the memory pressure intensifies, leading to further performance overhead.
2) \name~demonstrates significant performance benefits once the Meta table is established, as the number of iterations increases. It can be observed that, except for the initial iteration with higher overhead, the performance overhead decreases to 11\% of the 8-thread configuration after 10 iterations. Given that LLM training typically involves tens of thousands of iterations, the initialization phase is negligible. As a result, TensorTEE could commonly achieve comparable performance to SoftVN.

For the more complex GEMM workloads, TenAnalyzer efficiently detects and manages tensors by entries merging. Taking a 256*256 2D matrix with tiles of 64*64 as an example, due to the inherent multiple loop accesses in matrix multiplication, TenAnalyzer constructs tensor structures after a single GEMM operation, achieving a 98.8\% \textit{hit\_in} rate.

\textbf{Algorithms scalability limitations:} Our on-chip tensor management design optimizes the management of off-chip cachelines, making it applicable to algorithms of any scale. However, due to limited on-chip space, if an algorithm involves more than 512 tensors (number of MetaTable entries), the performance improvement gradually diminishes.

\subsection{Improvement of Delayed Verification}

\begin{figure}
    \centering
    \includegraphics[width=1\linewidth]{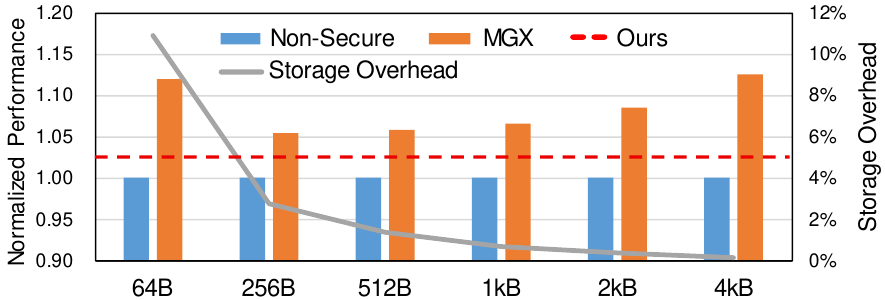}
    \caption{Delayed verification eliminates the computation stalls, achieving high performance and low storage overhead.}
    \label{fig:exp_npu}
\end{figure}

Figure~\ref{fig:exp_npu} illustrates the performance comparison and storage overhead of the NPU TEE at different MAC granularities. It can be observed that the storage overhead decreases as the MAC granularity increases. However, although there is a performance gain at smaller granularities due to fewer extra memory accesses, the overhead keeps increasing with larger granularity than 256B due to the stall problem.
The performance overhead reaches 13.0\% when MAC granularity is set to 4KB, which is unacceptable. In contrast, our proposed MAC tensor-based management reduces the storage overhead to negligible levels. Additionally, benefiting from delayed verification, the performance loss is only 2.5\%.

Since MAC regeneration and computation are executed in parallel, the overhead caused by the verification barriers refers to the overhead of MAC comparison. This operation can be completed in just a few cycles, making the overhead negligible.

\subsection{Improvement of Data Transfer Protocol}

\begin{figure}
    \centering
    \includegraphics[width=1\linewidth]{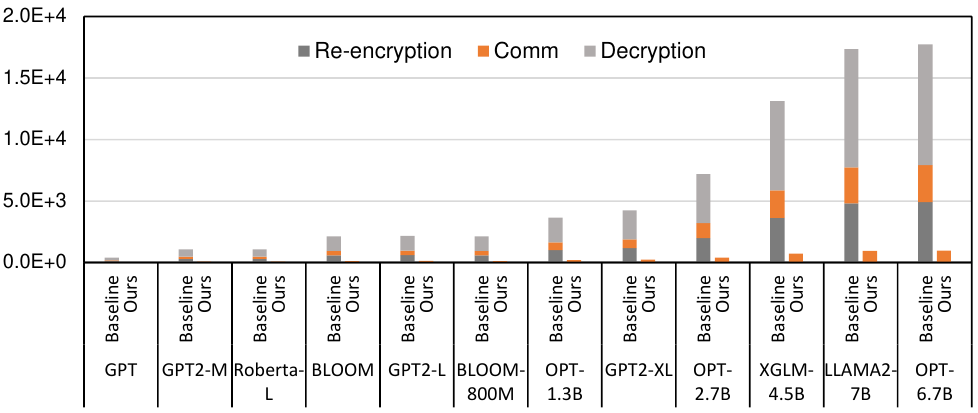}
    \caption{Breakdown for gradient transfer. \name~eliminates the re-encryption and decryption, and conceals data transfer overhead within the computation.}
    \label{fig:comm_breakdown}
\end{figure}

Figure~\ref{fig:comm_breakdown} presents the communication breakdown for gradient transfer. The following observations can be made from the figure:
1) Due to granularity mismatch, re-encryption and decryption are introduced before and after communication to align the granularities.
2) The presence of re-encryption and decryption prevents gradient transfer from being parallelized with NPU's gradient computation. Consequently, the communication time further increases.
\name~eliminates the need for re-encryption and decryption, allowing parallel execution of communication and computation. As a result, the communication performance is improved by 18.7x.

\subsection{Hardware Overhead}
We assume that the Meta Table consists of 512 entries to accommodate complex and dynamic applications. Each entry consists of an address range (64 bits for address and 92 bits for dimensions), a stride (10 bits), a VN (56 bits),  a MAC (56 bits), and flags (2 bits). Due to the limited number of tensors accessed simultaneously in common kernels like Attention (5 tensors) and Linear layer (3 tensors), the Tensor Filter requires only 10 entries, each containing 4 addresses along with VN and MAC. The on-chip bitmap cache needs only 6 KB to match the L3 cache. And the storage of poison bits needed for 512 tensor states is only 512 bits.  The combined storage requirement for the above components is only 24KB, resulting in an area of 0.0072$mm^{2}$ based on CACTI-7~\cite{balasubramonian2017cacti} simulation under 7 nm technology.

\section{Related Works}
\label{related_works}

\textbf{Privacy-preserving computing on LLM}:
Due to concerns about training data and model privacy, various security techniques have been proposed. MPC with garbled circuit and secret sharing~\cite{mo2023haac, dong2023puma, gupta2023sigma,hou2023ciphergpt,ding2023east, akimoto2023privformer}, excels in boolean logic circuits like ReLU computing but is not proficient in arithmetic logic such as matrix multiplication.  Differential privacy (DP)~\cite{dwork2014dp, hoory2021learning, li2021large, mireshghallah2022differentially,ponomareva2022training} adds noises like Gaussian noise into training data or gradients to improve its invisibility that the attacker cannot obtain the data information from differential inference. Although improving performance, DP degrades the model accuracy due to noise added. FHE~\cite{samardzic2021f1,hao2022iron,huang2022cheetah,raeini2023privacy, alchemist} is designed to compute with encrypted data or model, leading to unacceptable overhead due to the ciphertext expansion and heavy operations. TEE~\cite{costan2016sgx, southsecure, IntelTEE} is a secure region designed to protect the confidentiality and integrity of the data, model, and codes that are loaded into the protected memory region, achieved by memory encryption, Message Authentication Code (MAC) and Merkle Tree. TEE can provide strong privacy computing guarantees while incurring lower overhead compared to other privacy-preserving techniques, hence it has gained significant attention. Compared with VM-based TEEs, Merkle-Tree-based TEEs~\cite{feng2021penglai, costan2016sgx} like SGX offer enhanced security and remain competitive in high-security scenarios like blockchain~\cite{das2019fastkitten, cheng2019ekiden}, ML~\cite{kim2020vessels, zhang2021citadel}, and finance~\cite{sgx_china, sgx_story}.

\textbf{Memory encryption and integrity}:
Many works have been proposed to improve CPU TEE, including the counter-mode encryption~\cite{suh2003efficient}, optimizations to reduce the size of VNs~\cite{saileshwar2018morphable,yan2006improving}, meta-data caching~\cite{gassend2003caches, lee2016reducing}, and predicting VNs or using unverified VNs speculatively~\cite{lehman2016poisonivy, shi2006authentication, shi2005high}, scalability enhancement~\cite{feng2021penglai,feng2023mmt}. But all of them require version numbers in off-chip memory with large extra memory access overhead. The VN in SoftVN~\cite{umar2022softvn} is explicitly specified by software and compiled into instructions, thereby eliminating off-chip access. However, its static-dataflow assumption restricts its ability to adapt to dynamic and complex scenarios.
Recent efforts have extended the CPU TEE to the NPU TEE, involving the establishment of trusted communication channels~\cite{volos2018graviton, jang2019hix, CryptoMMU} and tensor-based VN management~\cite{hua2022mgx, lee2022tnpu, shrivastava2023securator, na2021common, hua2022guardnn}, predicting MACs~\cite{abdullah2023plutus}. However, these studies fail to address the compatibility issues at the granularity level with the CPU, resulting in significant communication overhead in collaborative computing.

\section{Conclusion}
\label{sec:conclusion}
On targeting efficient and secure collaborative heterogeneous computing, in this paper, we propose a unified tensor-wise secure memory management approach for the CPU and NPU with virtual support for CPU TEE, delayed verification for tensor-wise MAC, and direct data transfer protocols. These solutions enable efficient on-chip memory access and heterogeneous data transfer in LLM collaborative computing, ultimately enhancing end-to-end performance.

\section{Acknowledgements}
We extend our sincere gratitude to our shepherd, Boris Pismenny, and the anonymous reviewers for their invaluable feedback and suggestions, which greatly contributed to the improvement of our paper. This work is partially supported by the National Key R\&D Program of China (under Grant 2023YFB4502200), the NSF of China (under Grants U22A2028, 61925208, 62222214, 62341411, 62302478, 62372436, 62302482, 62102398, 62102399, U20A20227, 62302483, 62302480), Strategic Priority Research Program of the Chinese Academy of Sciences, (Grant No. XDB0660300, XDB0660301, XDB0660302), CAS Project for Young Scientists in Basic Research (YSBR-029), Youth Innovation Promotion Association CAS and Xplore Prize.


\bibliographystyle{plain}
\bibliography{references}

\begin{thebibliography}{10}

\bibitem{amdsev}
Amd secure encrypted virtualization (sev) - amd.
\newblock \url{https://developer.amd.com/sev/}, 2019.

\bibitem{abdullah2023plutus}
Rahaf Abdullah, Huiyang Zhou, and Amro Awad.
\newblock Plutus: Bandwidth-efficient memory security for gpus.
\newblock In {\em 2023 IEEE International Symposium on High-Performance Computer Architecture (HPCA)}, pages 543--555. IEEE, 2023.

\bibitem{akimoto2023privformer}
Yoshimasa Akimoto, Kazuto Fukuchi, Youhei Akimoto, and Jun Sakuma.
\newblock Privformer: Privacy-preserving transformer with mpc.
\newblock In {\em 2023 IEEE 8th European Symposium on Security and Privacy (EuroS\&P)}, pages 392--410. IEEE, 2023.

\bibitem{CryptoMMU}
Faiz Alam, Hyokeun Lee, Abhishek Bhattacharjee, and Amro Awad.
\newblock Cryptommu: Enabling scalable and secure access control of third-party accelerators.
\newblock In {\em 2023 56th Annual IEEE/ACM International Symposium on Microarchitecture (MICRO)}. IEEE, 2023.

\bibitem{MI300}
AMD.
\newblock Amd instinct mi300 series accelerators.
\newblock \url{www.amd.com/en/products/accelerators/instinct/mi300/mi300a.html}.

\bibitem{CCA}
ARM.
\newblock Arm cca security model 1.0.
\newblock {\em whilte paper}, 2021.

\bibitem{balasubramonian2017cacti}
Rajeev Balasubramonian, Andrew~B Kahng, Naveen Muralimanohar, Ali Shafiee, and Vaishnav Srinivas.
\newblock Cacti 7: New tools for interconnect exploration in innovative off-chip memories.
\newblock {\em ACM Transactions on Architecture and Code Optimization (TACO)}, 14(2):1--25, 2017.

\bibitem{binkert2011gem5}
Nathan Binkert, Bradford Beckmann, Gabriel Black, Steven~K Reinhardt, Ali Saidi, Arkaprava Basu, Joel Hestness, Derek~R Hower, Tushar Krishna, Somayeh Sardashti, et~al.
\newblock The gem5 simulator.
\newblock {\em ACM SIGARCH computer architecture news}, 39(2):1--7, 2011.

\bibitem{chakraborty2018adversarial}
Anirban Chakraborty, Manaar Alam, Vishal Dey, Anupam Chattopadhyay, and Debdeep Mukhopadhyay.
\newblock Adversarial attacks and defences: A survey.
\newblock {\em arXiv preprint arXiv:1810.00069}, 2018.

\bibitem{hmtt_project}
Mingyu Chen and Yungang Bao.
\newblock Hmtt: Hybrid memory trace toolkit.
\newblock {\em http://asg.ict.ac.cn/hmtt/}, 2019.

\bibitem{cheng2019ekiden}
Raymond Cheng, Fan Zhang, Jernej Kos, Warren He, Nicholas Hynes, Noah Johnson, Ari Juels, Andrew Miller, and Dawn Song.
\newblock Ekiden: A platform for confidentiality-preserving, trustworthy, and performant smart contracts.
\newblock In {\em 2019 IEEE European Symposium on Security and Privacy (EuroS\&P)}, pages 185--200. IEEE, 2019.

\bibitem{choukse2020buddy}
Esha Choukse, Michael~B Sullivan, Mike O’Connor, Mattan Erez, Jeff Pool, David Nellans, and Stephen~W Keckler.
\newblock Buddy compression: Enabling larger memory for deep learning and hpc workloads on gpus.
\newblock In {\em 2020 ACM/IEEE 47th Annual International Symposium on Computer Architecture (ISCA)}, pages 926--939. IEEE, 2020.

\bibitem{costan2016sgx}
Victor Costan and Srinivas Devadas.
\newblock Intel sgx explained.
\newblock {\em Cryptology ePrint Archive}, 2016.

\bibitem{das2019fastkitten}
Poulami Das, Lisa Eckey, Tommaso Frassetto, David Gens, Kristina Host{\'a}kov{\'a}, Patrick Jauernig, Sebastian Faust, and Ahmad-Reza Sadeghi.
\newblock $\{$FastKitten$\}$: Practical smart contracts on bitcoin.
\newblock In {\em 28th USENIX Security Symposium (USENIX Security 19)}, pages 801--818, 2019.

\bibitem{deng2022strongbox}
Yunjie Deng, Chenxu Wang, Shunchang Yu, Shiqing Liu, Zhenyu Ning, Kevin Leach, Jin Li, Shoumeng Yan, Zhengyu He, Jiannong Cao, et~al.
\newblock Strongbox: A gpu tee on arm endpoints.
\newblock In {\em Proceedings of the 2022 ACM SIGSAC Conference on Computer and Communications Security}, pages 769--783, 2022.

\bibitem{devices2006amd64}
A~Micro Devices.
\newblock Amd64 architecture programmer’s manual volume 2: System programming.
\newblock {\em 2006}, 2006.

\bibitem{ding2023east}
Yuanchao Ding, Hua Guo, Yewei Guan, Weixin Liu, Jiarong Huo, Zhenyu Guan, and Xiyong Zhang.
\newblock East: Efficient and accurate secure transformer framework for inference.
\newblock {\em arXiv preprint arXiv:2308.09923}, 2023.

\bibitem{dong2023puma}
Ye~Dong, Wen-jie Lu, Yancheng Zheng, Haoqi Wu, Derun Zhao, Jin Tan, Zhicong Huang, Cheng Hong, Tao Wei, and Wenguang Cheng.
\newblock Puma: Secure inference of llama-7b in five minutes.
\newblock {\em arXiv preprint arXiv:2307.12533}, 2023.

\bibitem{dwork2014dp}
Cynthia Dwork, Aaron Roth, et~al.
\newblock The algorithmic foundations of differential privacy.
\newblock {\em Foundations and Trends{\textregistered} in Theoretical Computer Science}, 9(3--4):211--407, 2014.

\bibitem{GH200}
Ashraf Eassa, Ashwin Nanjappa, Jinho Suh, Yiheng Zhang, and Zhihan Jiang.
\newblock Leading mlperf inference v3.1 results with nvidia gh200 grace hopper superchip debut.
\newblock \url{https://developer.nvidia.com/blog/leading-mlperf-inference-v3-1-results-gh200-grace-hopper-superchip-debut/}.

\bibitem{feng2023mmt}
Erhu Feng, Dong Du, Yubin Xia, and Haibo Chen.
\newblock Efficient distributed secure memory with migratable merkle tree.
\newblock In {\em 2023 IEEE International Symposium on High-Performance Computer Architecture (HPCA)}, pages 347--360. IEEE, 2023.

\bibitem{feng2021penglai}
Erhu Feng, Xu~Lu, Dong Du, Bicheng Yang, Xueqiang Jiang, Yubin Xia, Binyu Zang, and Haibo Chen.
\newblock Scalable memory protection in the penglai enclave.
\newblock In {\em 15th USENIX Symposium on Operating Systems Design and Implementation (OSDI 21)}, pages 275--294, 2021.

\bibitem{gassend2003caches}
Blaise Gassend, G~Edward Suh, Dwaine Clarke, Marten Van~Dijk, and Srinivas Devadas.
\newblock Caches and hash trees for efficient memory integrity verification.
\newblock In {\em The Ninth International Symposium on High-Performance Computer Architecture, 2003. HPCA-9 2003. Proceedings.}, pages 295--306. IEEE, 2003.

\bibitem{gueron2016mee}
Shay Gueron.
\newblock A memory encryption engine suitable for general purpose processors.(2016), 2016.

\bibitem{gueron2016memory}
Shay Gueron.
\newblock Memory encryption for general-purpose processors.
\newblock {\em IEEE Security \& Privacy}, 14(6):54--62, 2016.

\bibitem{gupta2023sigma}
Kanav Gupta, Neha Jawalkar, Ananta Mukherjee, Nishanth Chandran, Divya Gupta, Ashish Panwar, and Rahul Sharma.
\newblock Sigma: Secure gpt inference with function secret sharing.
\newblock {\em Cryptology ePrint Archive}, 2023.

\bibitem{han2023real}
Husheng Han, Xing Hu, Yifan Hao, Kaidi Xu, Pucheng Dang, Ying Wang, Yongwei Zhao, Zidong Du, Qi~Guo, Yanzhi Wang, et~al.
\newblock Real-time robust video object detection system against physical-world adversarial attacks.
\newblock {\em IEEE Transactions on Computer-Aided Design of Integrated Circuits and Systems}, 2023.

\bibitem{han2021scalecert}
Husheng Han, Kaidi Xu, Xing Hu, Xiaobing Chen, Ling Liang, Zidong Du, Qi~Guo, Yanzhi Wang, and Yunji Chen.
\newblock Scalecert: Scalable certified defense against adversarial patches with sparse superficial layers.
\newblock {\em Advances in Neural Information Processing Systems}, 34:28169--28181, 2021.

\bibitem{hao2022iron}
Meng Hao, Hongwei Li, Hanxiao Chen, Pengzhi Xing, Guowen Xu, and Tianwei Zhang.
\newblock Iron: Private inference on transformers.
\newblock {\em Advances in Neural Information Processing Systems}, 35:15718--15731, 2022.

\bibitem{hildebrand2020autotm}
Mark Hildebrand, Jawad Khan, Sanjeev Trika, Jason Lowe-Power, and Venkatesh Akella.
\newblock Autotm: Automatic tensor movement in heterogeneous memory systems using integer linear programming.
\newblock In {\em Proceedings of the Twenty-Fifth International Conference on Architectural Support for Programming Languages and Operating Systems}, pages 875--890, 2020.

\bibitem{hoory2021learning}
Shlomo Hoory, Amir Feder, Avichai Tendler, Sofia Erell, Alon Peled-Cohen, Itay Laish, Hootan Nakhost, Uri Stemmer, Ayelet Benjamini, Avinatan Hassidim, et~al.
\newblock Learning and evaluating a differentially private pre-trained language model.
\newblock In {\em Findings of the Association for Computational Linguistics: EMNLP 2021}, pages 1178--1189, 2021.

\bibitem{hou2023ciphergpt}
Xiaoyang Hou, Jian Liu, Jingyu Li, Yuhan Li, Wen-jie Lu, Cheng Hong, and Kui Ren.
\newblock Ciphergpt: Secure two-party gpt inference.
\newblock {\em Cryptology ePrint Archive}, 2023.

\bibitem{hua2022guardnn}
Weizhe Hua, Muhammad Umar, Zhiru Zhang, and G~Edward Suh.
\newblock Guardnn: secure accelerator architecture for privacy-preserving deep learning.
\newblock In {\em Proceedings of the 59th ACM/IEEE Design Automation Conference}, pages 349--354, 2022.

\bibitem{hua2022mgx}
Weizhe Hua, Muhammad Umar, Zhiru Zhang, and G~Edward Suh.
\newblock Mgx: Near-zero overhead memory protection for data-intensive accelerators.
\newblock In {\em Proceedings of the 49th Annual International Symposium on Computer Architecture}, pages 726--741, 2022.

\bibitem{huang2002keeping}
Andrew Huang.
\newblock Keeping secrets in hardware: The microsoft xboxtm case study.
\newblock In {\em International Workshop on Cryptographic Hardware and Embedded Systems}, pages 213--227. Springer, 2002.

\bibitem{huang2020swapadvisor}
Chien-Chin Huang, Gu~Jin, and Jinyang Li.
\newblock Swapadvisor: Pushing deep learning beyond the gpu memory limit via smart swapping.
\newblock In {\em Proceedings of the Twenty-Fifth International Conference on Architectural Support for Programming Languages and Operating Systems}, pages 1341--1355, 2020.

\bibitem{huang2014hmtt}
Yongbing Huang, Licheng Chen, Zehan Cui, Yuan Ruan, Yungang Bao, Mingyu Chen, and Ninghui Sun.
\newblock Hmtt: A hybrid hardware/software tracing system for bridging the dram access trace's semantic gap.
\newblock {\em ACM Transactions on Architecture and Code Optimization (TACO)}, 11(1):1--25, 2014.

\bibitem{huang2022cheetah}
Zhicong Huang, Wen-jie Lu, Cheng Hong, and Jiansheng Ding.
\newblock Cheetah: Lean and fast secure two-party deep neural network inference.
\newblock In {\em 31st USENIX Security Symposium (USENIX Security 22)}, pages 809--826, 2022.

\bibitem{hunt2020telekine}
Tyler Hunt, Zhipeng Jia, Vance Miller, Ariel Szekely, Yige Hu, Christopher~J Rossbach, and Emmett Witchel.
\newblock Telekine: Secure computing with cloud gpus.
\newblock In {\em 17th USENIX Symposium on Networked Systems Design and Implementation (NSDI 20)}, pages 817--833, 2020.

\bibitem{sgx_china}
Intel.
\newblock Intel china financial privacy computing gallery.
\newblock \url{https://www.intel.cn/content/www/cn/zh/artificial-intelligence/china-financial-privacy-computing-gallery.html}.

\bibitem{sgx_story}
Intel.
\newblock Sgx bank customer story.
\newblock \url{https://www.intel.com/content/www/us/en/customer-spotlight/stories/eperi-sgx-customer-story.html}.

\bibitem{TDX}
Intel.
\newblock Intel trust domain extensions.
\newblock {\em white paper}, 2023.

\bibitem{jang2019hix}
Insu Jang, Adrian Tang, Taehoon Kim, Simha Sethumadhavan, and Jaehyuk Huh.
\newblock Heterogeneous isolated execution for commodity gpus.
\newblock In {\em Proceedings of the Twenty-Fourth International Conference on Architectural Support for Programming Languages and Operating Systems}, pages 455--468, 2019.

\bibitem{jin2018layer}
Hai Jin, Bo~Liu, Wenbin Jiang, Yang Ma, Xuanhua Shi, Bingsheng He, and Shaofeng Zhao.
\newblock Layer-centric memory reuse and data migration for extreme-scale deep learning on many-core architectures.
\newblock {\em ACM Transactions on Architecture and Code Optimization (TACO)}, 15(3):1--26, 2018.

\bibitem{jouppi2017tpu_v1}
Norman~P Jouppi, Cliff Young, Nishant Patil, David Patterson, Gaurav Agrawal, Raminder Bajwa, Sarah Bates, Suresh Bhatia, Nan Boden, Al~Borchers, et~al.
\newblock In-datacenter performance analysis of a tensor processing unit.
\newblock In {\em Proceedings of the 44th annual international symposium on computer architecture}, pages 1--12, 2017.

\bibitem{AlphaFold2021}
John Jumper, Richard Evans, Alexander Pritzel, Tim Green, Michael Figurnov, Olaf Ronneberger, Kathryn Tunyasuvunakool, Russ Bates, Augustin {\v{Z}}{\'\i}dek, Anna Potapenko, Alex Bridgland, Clemens Meyer, Simon A~A Kohl, Andrew~J Ballard, Andrew Cowie, Bernardino Romera-Paredes, Stanislav Nikolov, Rishub Jain, Jonas Adler, Trevor Back, Stig Petersen, David Reiman, Ellen Clancy, Michal Zielinski, Martin Steinegger, Michalina Pacholska, Tamas Berghammer, Sebastian Bodenstein, David Silver, Oriol Vinyals, Andrew~W Senior, Koray Kavukcuoglu, Pushmeet Kohli, and Demis Hassabis.
\newblock Highly accurate protein structure prediction with {AlphaFold}.
\newblock {\em Nature}, 596(7873):583--589, 2021.

\bibitem{kadam2018rcoal}
Gurunath Kadam, Danfeng Zhang, and Adwait Jog.
\newblock Rcoal: mitigating gpu timing attack via subwarp-based randomized coalescing techniques.
\newblock In {\em 2018 IEEE international symposium on high performance computer architecture (HPCA)}, pages 156--167. IEEE, 2018.

\bibitem{kaplan2016amd}
David Kaplan, Jeremy Powell, and Tom Woller.
\newblock Amd memory encryption.
\newblock {\em White paper}, 2021.

\bibitem{karimi2018timing}
Elmira Karimi, Zhen~Hang Jiang, Yunsi Fei, and David Kaeli.
\newblock A timing side-channel attack on a mobile gpu.
\newblock In {\em 2018 IEEE 36th International Conference on Computer Design (ICCD)}, pages 67--74. IEEE, 2018.

\bibitem{kim2020vessels}
Kyungtae Kim, Chung~Hwan Kim, Junghwan"~John" Rhee, Xiao Yu, Haifeng Chen, Dave Tian, and Byoungyoung Lee.
\newblock Vessels: Efficient and scalable deep learning prediction on trusted processors.
\newblock In {\em Proceedings of the 11th ACM Symposium on Cloud Computing}, pages 462--476, 2020.

\bibitem{kim2015ramulator}
Yoongu Kim, Weikun Yang, and Onur Mutlu.
\newblock Ramulator: A fast and extensible dram simulator.
\newblock {\em IEEE Computer architecture letters}, 15(1):45--49, 2015.

\bibitem{lee2020off}
Dayeol Lee, Dongha Jung, Ian~T Fang, Chia-Che Tsai, and Raluca~Ada Popa.
\newblock An off-chip attack on hardware enclaves via the memory bus.
\newblock In {\em 29th USENIX Security Symposium (USENIX Security 20)}, 2020.

\bibitem{lee2020keystone}
Dayeol Lee, David Kohlbrenner, Shweta Shinde, Krste Asanovi{\'c}, and Dawn Song.
\newblock Keystone: An open framework for architecting trusted execution environments.
\newblock In {\em Proceedings of the Fifteenth European Conference on Computer Systems}, pages 1--16, 2020.

\bibitem{lee2016reducing}
Junghoon Lee, Taehoon Kim, and Jaehyuk Huh.
\newblock Reducing the memory bandwidth overheads of hardware security support for multi-core processors.
\newblock {\em IEEE Transactions on Computers}, 65(11):3384--3397, 2016.

\bibitem{lee2023secureloop}
Kyungmi Lee, Mengjia Yan, Joel Emer, and Anantha Chandrakasan.
\newblock Secureloop: Design space exploration of secure dnn accelerators.
\newblock In {\em Proceedings of the 56th Annual IEEE/ACM International Symposium on Microarchitecture}, pages 194--208, 2023.

\bibitem{lee2022tnpu}
Sunho Lee, Jungwoo Kim, Seonjin Na, Jongse Park, and Jaehyuk Huh.
\newblock Tnpu: Supporting trusted execution with tree-less integrity protection for neural processing unit.
\newblock In {\em 2022 IEEE International Symposium on High-Performance Computer Architecture (HPCA)}, pages 229--243. IEEE, 2022.

\bibitem{lehman2016poisonivy}
Tamara~Silbergleit Lehman, Andrew~D Hilton, and Benjamin~C Lee.
\newblock Poisonivy: Safe speculation for secure memory.
\newblock In {\em 2016 49th Annual IEEE/ACM International Symposium on Microarchitecture (MICRO)}, pages 1--13. IEEE, 2016.

\bibitem{li2021large}
Xuechen Li, Florian Tramer, Percy Liang, and Tatsunori Hashimoto.
\newblock Large language models can be strong differentially private learners.
\newblock {\em arXiv preprint arXiv:2110.05679}, 2021.

\bibitem{mireshghallah2022differentially}
Fatemehsadat Mireshghallah, Arturs Backurs, Huseyin~A Inan, Lukas Wutschitz, and Janardhan Kulkarni.
\newblock Differentially private model compression.
\newblock {\em Advances in Neural Information Processing Systems}, 35:29468--29483, 2022.

\bibitem{mo2023haac}
Jianqiao Mo, Jayanth Gopinath, and Brandon Reagen.
\newblock Haac: A hardware-software co-design to accelerate garbled circuits.
\newblock In {\em Proceedings of the 50th Annual International Symposium on Computer Architecture}, pages 1--13, 2023.

\bibitem{alchemist}
Jianan Mu, Husheng Han, Shangyi Shi, Jing Ye, Zizhen Liu, Shengwen Liang, Meng Li, Mingzhe Zhang, Song Bian, Xing Hu, Huaiwei Li, and Xiaowei Li.
\newblock Alchemist: A unified accelerator architecture for cross-scheme fully homomorphic encryption.
\newblock In {\em 61th Design Automation Conference (DAC)}, 2024.

\bibitem{na2021common}
Seonjin Na, Sunho Lee, Yeonjae Kim, Jongse Park, and Jaehyuk Huh.
\newblock Common counters: Compressed encryption counters for secure gpu memory.
\newblock In {\em 2021 IEEE International Symposium on High-Performance Computer Architecture (HPCA)}, pages 1--13. IEEE, 2021.

\bibitem{naghibijouybari2018rendered}
Hoda Naghibijouybari, Ajaya Neupane, Zhiyun Qian, and Nael Abu-Ghazaleh.
\newblock Rendered insecure: Gpu side channel attacks are practical.
\newblock In {\em Proceedings of the 2018 ACM SIGSAC conference on computer and communications security}, pages 2139--2153, 2018.

\bibitem{peng2020capuchin}
Xuan Peng, Xuanhua Shi, Hulin Dai, Hai Jin, Weiliang Ma, Qian Xiong, Fan Yang, and Xuehai Qian.
\newblock Capuchin: Tensor-based gpu memory management for deep learning.
\newblock In {\em Proceedings of the Twenty-Fifth International Conference on Architectural Support for Programming Languages and Operating Systems}, pages 891--905, 2020.

\bibitem{ponomareva2022training}
Natalia Ponomareva, Jasmijn Bastings, and Sergei Vassilvitskii.
\newblock Training text-to-text transformers with privacy guarantees.
\newblock In {\em Findings of the Association for Computational Linguistics: ACL 2022}, pages 2182--2193, 2022.

\bibitem{raeini2023privacy}
Mohammad Raeini.
\newblock Privacy-preserving large language models (ppllms).
\newblock {\em Available at SSRN 4512071}, 2023.

\bibitem{rajbhandari2020zero}
Samyam Rajbhandari, Jeff Rasley, Olatunji Ruwase, and Yuxiong He.
\newblock Zero: Memory optimizations toward training trillion parameter models.
\newblock In {\em SC20: International Conference for High Performance Computing, Networking, Storage and Analysis}, pages 1--16. IEEE, 2020.

\bibitem{rasley2020deepspeed}
Jeff Rasley, Samyam Rajbhandari, Olatunji Ruwase, and Yuxiong He.
\newblock Deepspeed: System optimizations enable training deep learning models with over 100 billion parameters.
\newblock In {\em Proceedings of the 26th ACM SIGKDD International Conference on Knowledge Discovery \& Data Mining}, pages 3505--3506, 2020.

\bibitem{ren2021sentinel}
Jie Ren, Jiaolin Luo, Kai Wu, Minjia Zhang, Hyeran Jeon, and Dong Li.
\newblock Sentinel: Efficient tensor migration and allocation on heterogeneous memory systems for deep learning.
\newblock In {\em 2021 IEEE International Symposium on High-Performance Computer Architecture (HPCA)}, pages 598--611. IEEE, 2021.

\bibitem{ren2021zero}
Jie Ren, Samyam Rajbhandari, Reza~Yazdani Aminabadi, Olatunji Ruwase, Shuangyan Yang, Minjia Zhang, Dong Li, and Yuxiong He.
\newblock Zero-offload: Democratizing billion-scale model training.
\newblock In {\em 2021 USENIX Annual Technical Conference (USENIX ATC 21)}, pages 551--564, 2021.

\bibitem{rhu2016vdnn}
Minsoo Rhu, Natalia Gimelshein, Jason Clemons, Arslan Zulfiqar, and Stephen~W Keckler.
\newblock vdnn: Virtualized deep neural networks for scalable, memory-efficient neural network design.
\newblock In {\em 2016 49th Annual IEEE/ACM International Symposium on Microarchitecture (MICRO)}, pages 1--13. IEEE, 2016.

\bibitem{rogers2007BMT}
Brian Rogers, Siddhartha Chhabra, Milos Prvulovic, and Yan Solihin.
\newblock Using address independent seed encryption and bonsai merkle trees to make secure processors os-and performance-friendly.
\newblock In {\em 40th Annual IEEE/ACM International Symposium on Microarchitecture (MICRO 2007)}, pages 183--196. IEEE, 2007.

\bibitem{saileshwar2018morphable}
Gururaj Saileshwar, Prashant~J Nair, Prakash Ramrakhyani, Wendy Elsasser, Jose~A Joao, and Moinuddin~K Qureshi.
\newblock Morphable counters: Enabling compact integrity trees for low-overhead secure memories.
\newblock In {\em 2018 51st Annual IEEE/ACM International Symposium on Microarchitecture (MICRO)}, pages 416--427. IEEE, 2018.

\bibitem{samardzic2021f1}
Nikola Samardzic, Axel Feldmann, Aleksandar Krastev, Srinivas Devadas, Ronald Dreslinski, Christopher Peikert, and Daniel Sanchez.
\newblock F1: A fast and programmable accelerator for fully homomorphic encryption.
\newblock In {\em MICRO-54: 54th Annual IEEE/ACM International Symposium on Microarchitecture}, pages 238--252, 2021.

\bibitem{shi2005high}
Weidong Shi, H-h~S Lee, Mrinmoy Ghosh, Chenghuai Lu, and Alexandra Boldyreva.
\newblock High efficiency counter mode security architecture via prediction and precomputation.
\newblock In {\em 32nd International Symposium on Computer Architecture (ISCA'05)}, pages 14--24. IEEE, 2005.

\bibitem{shi2006authentication}
Weidong Shi and Hsien-Hsin~S Lee.
\newblock Authentication control point and its implications for secure processor design.
\newblock In {\em 2006 39th Annual IEEE/ACM International Symposium on Microarchitecture (MICRO'06)}, pages 103--112. IEEE, 2006.

\bibitem{shi2005towards}
Weidong Shi, Hsien-Hsin~S Lee, Chenghuai Lu, and Mrinmoy Ghosh.
\newblock Towards the issues in architectural support for protection of software execution.
\newblock {\em ACM SIGARCH Computer Architecture News}, 33(1):6--15, 2005.

\bibitem{shrivastava2023securator}
Nivedita Shrivastava and Smruti~Ranjan Sarangi.
\newblock Securator: A fast and secure neural processing unit.
\newblock In {\em 2023 IEEE International Symposium on High-Performance Computer Architecture (HPCA)}, pages 1127--1139. IEEE, 2023.

\bibitem{southsecure}
Tobin South, Guy Zuskind, Robert Mahari, and Thomas Hardjono.
\newblock Secure community transformers: Private pooled data for llms.
\newblock \url{https://transformers.mit.edu/SecureCommunityTransfomersMITSouth.pdf}.

\bibitem{stefanov2018pathoram}
Emil Stefanov, Marten~van Dijk, Elaine Shi, T-H~Hubert Chan, Christopher Fletcher, Ling Ren, Xiangyao Yu, and Srinivas Devadas.
\newblock Path oram: an extremely simple oblivious ram protocol.
\newblock {\em Journal of the ACM (JACM)}, 65(4):1--26, 2018.

\bibitem{suh2003efficient}
G~Edward Suh, Dwaine Clarke, Blaise Gasend, Marten Van~Dijk, and Srinivas Devadas.
\newblock Efficient memory integrity verification and encryption for secure processors.
\newblock In {\em Proceedings. 36th Annual IEEE/ACM International Symposium on Microarchitecture, 2003. MICRO-36.}, pages 339--350. IEEE, 2003.

\bibitem{tesla}
Tesla.
\newblock Tesla hardware news.
\newblock \url{https://www.autopilotreview.com/tesla-hardware-4-rolling-out-to-new-vehicles/}, 2023.

\bibitem{umar2022softvn}
Muhammad Umar, Weizhe Hua, Zhiru Zhang, and G~Edward Suh.
\newblock Softvn: Efficient memory protection via software-provided version numbers.
\newblock In {\em Proceedings of the 49th Annual International Symposium on Computer Architecture}, pages 160--172, 2022.

\bibitem{volos2018graviton}
Stavros Volos, Kapil Vaswani, and Rodrigo Bruno.
\newblock Graviton: Trusted execution environments on gpus.
\newblock In {\em 13th USENIX Symposium on Operating Systems Design and Implementation (OSDI 18)}, pages 681--696, 2018.

\bibitem{wegman1981new}
Mark~N Wegman and J~Lawrence Carter.
\newblock New hash functions and their use in authentication and set equality.
\newblock {\em Journal of computer and system sciences}, 22(3):265--279, 1981.

\bibitem{yan2006improving}
Chenyu Yan, Daniel Englender, Milos Prvulovic, Brian Rogers, and Yan Solihin.
\newblock Improving cost, performance, and security of memory encryption and authentication.
\newblock {\em ACM SIGARCH Computer Architecture News}, 34(2):179--190, 2006.

\bibitem{IntelTEE}
Wei Yu, Zhiqiang Li, Qingqing Li, and Qiyuan Long.
\newblock Effectively address large language model privacy and security challenges with a trusted execution environment based on intel sgx.
\newblock \url{https://www.intel.cn/content/www/cn/zh/customer-spotlight/cases/privacy-security-challenge-large-language-model.html}.

\bibitem{yuan2022adaptive}
Shougang Yuan, Amro Awad, Ardhi Wiratama~Baskara Yudha, Yan Solihin, and Huiyang Zhou.
\newblock Adaptive security support for heterogeneous memory on gpus.
\newblock In {\em 2022 IEEE International Symposium on High-Performance Computer Architecture (HPCA)}, pages 213--228. IEEE, 2022.

\bibitem{zhang2021citadel}
Chengliang Zhang, Junzhe Xia, Baichen Yang, Huancheng Puyang, Wei Wang, Ruichuan Chen, Istemi~Ekin Akkus, Paarijaat Aditya, and Feng Yan.
\newblock Citadel: Protecting data privacy and model confidentiality for collaborative learning.
\newblock In {\em Proceedings of the ACM Symposium on Cloud Computing}, pages 546--561, 2021.

\bibitem{zuo2021sealing}
Pengfei Zuo, Yu~Hua, Ling Liang, Xinfeng Xie, Xing Hu, and Yuan Xie.
\newblock Sealing neural network models in encrypted deep learning accelerators.
\newblock In {\em 2021 58th ACM/IEEE Design Automation Conference (DAC)}, pages 1255--1260. IEEE, 2021.

\end{thebibliography}

\end{document}